\documentclass[twoside,twocolumn,english]{revtex4-1}
\usepackage[T1]{fontenc}
\usepackage[latin9]{inputenc}
\usepackage{textcomp}
\usepackage{amsmath}
\usepackage{amssymb}
\usepackage{graphicx}
\usepackage{wasysym}

\makeatletter

\newcommand{\noun}[1]{\textsc{#1}}

\newenvironment{lyxlist}[1]
	{\begin{list}{}
		{\settowidth{\labelwidth}{#1}
		 \setlength{\leftmargin}{\labelwidth}
		 \addtolength{\leftmargin}{\labelsep}
		 }}
	{\end{list}}

\usepackage{soul}
\usepackage{color}
\usepackage{babel}

\makeatother

\usepackage{babel}
\begin{document}

\title{Complexity Bounds on Quantum Search Algorithms in finite-dimensional
Networks}

\author{Stefan Boettcher$^{1,*}$, Shanshan Li$^{1}$, Tharso D. Fernandes$^{2,3}$,
and Renato Portugal$^{2}$}

\affiliation{$^{1}$Department of Physics, Emory University, Atlanta, GA 30322;
USA ~~\\
 $^{2}$Laborat{ó}rio Nacional de Computa{ç}{ã}o Cient{í}fica,
Petrópolis, RJ 25651-075; Brazil~~\\
 $^{3}$Universidade Federal do Espírito Santo, Alegre, ES 29500-000;
Brazil}
\begin{abstract}
We establish a lower bound concerning the computational complexity
of Grover's algorithms on fractal networks. This bound provides general
predictions for the quantum advantage gained for searching unstructured
lists. It yields a fundamental criterion, derived from quantum transport
properties, for the improvement a quantum search algorithm achieves
over the corresponding classical search in a network based solely
on its spectral dimension, $d_{s}$. Our analysis employs recent advances
in the interpretation of the venerable real-space renormalization
group (RG) as applied to quantum walks. It clarifies the competition
between Grover's abstract algorithm, i.e., a rotation in Hilbert space,
and quantum transport in an actual geometry. The latter is characterized
in terms of the quantum walk dimension $d_{w}^{Q}$ and the spatial
(fractal) dimension $d_{f}$ that is summarized simply by the spectral
dimension of the network. The analysis simultaneously determines the
optimal time for a quantum measurement and the probability for successfully
pin-pointing a marked element in the network. The RG further encompasses
an optimization scheme devised by Tulsi that allows to tune this probability
to certainty, leaving quantum transport as the only limiting process.
It considers entire families of problems to be studied, thereby establishing
large universality classes for quantum search, which we verify with
extensive simulations. The methods we develop could point the way
towards systematic studies of universality classes in computational
complexity to enable modification and control of search behavior. 
\end{abstract}
\maketitle

\section{Introduction\label{sec:Intro}}

\begin{figure}[b]
\hfill{}\includegraphics[viewport=280bp 270bp 575bp 430bp,clip,width=0.85\columnwidth]{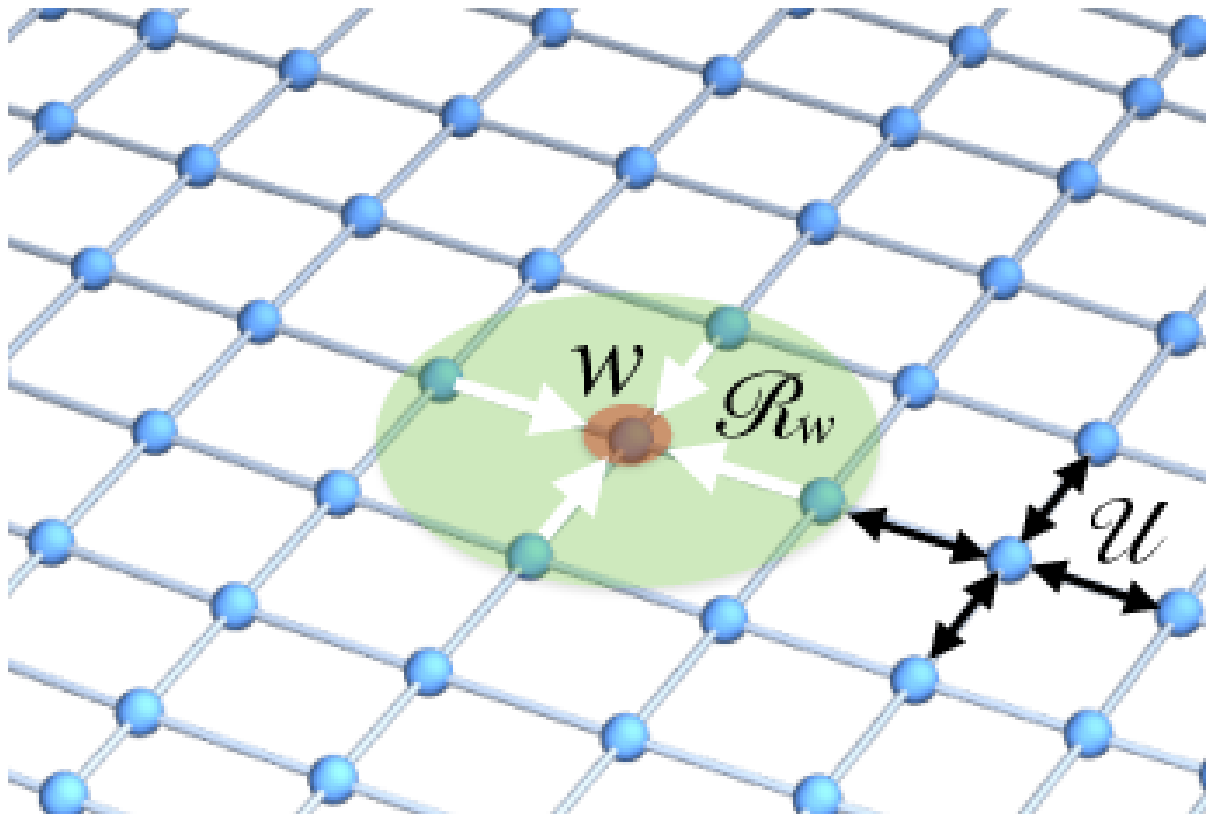}\hfill{}

\caption{\label{fig:MarkedSite}In Grover's quantum search algorithm, the search-operator
${\cal R}_{w}$ affects an accumulation (white arrows) of the wave-function
$\psi_{x,t}$ onto the marked site, $w$, but only from its neighborhood
(shaded). The walk-operator ${\cal U}$ transports $\psi$ uniformly
between mutually linked sites (black arrows), replenishing neighbors
of $w$ in the process. This distinction only arises in a finite-dimensional
geometry; it is moot in Grover's original work\ \cite{Gro97a}, where
all $N$ sites are linked.}
\end{figure}

Quantum walk present one of the frameworks in which quantum computing
can satisfy its promise to provide a speed-up over classical computation.
It applies to a significant number of interesting problems such as
quantum search~\cite{Gro97a,SKW03}, element distinctness~\cite{Ambainis07,ambainis_2003a},
graph isomorphisms~\cite{Shiau05,Douglas2008,Rudinger13}, and circuit
design~\cite{Qiang2016}. Analog to random walks, which have been
of fundamental importance for the development of stochastic algorithms
in classical computing~\cite{MM11,Motwani95}, quantum walks have
been established as a universal model of quantum computing~\cite{Childs09,lovett_2010a,Childs13}.
Similarly, the \emph{physical} properties of quantum walks in localization~\cite{inui_2004a,Crespi13,Vakulchyk17},
in entanglement~\cite{Omar06,carneiro_2005a,Schreiber12}, in interference~\cite{Peruzzo1500},
in decoherence~\cite{schreiber_2011a}, in topological invariants~\cite{Ramasesh17},
etc~\cite{VA12}, rival classical diffusion as an important transport
problem~\cite{Metzler04,Weiss94,Redner01}. In fact, numerous experimental
realizations of quantum walks have been proposed and studied in waveguides~\cite{Perets08},
in photonics~\cite{Schreiber12,Sansoni12,Crespi13,Qiang2016}, and
in atomic physics~\cite{Figgatt2017,Weitenberg11,Eckert05,Travaglione02}.
Photosynthesis provides even a natural occurrence~\cite{Engel07,Mohseni2008}.

\begin{figure}
\vspace{-0.2cm}

\hfill{}\includegraphics[viewport=20bp 0bp 720bp 550bp,clip,width=0.95\columnwidth]{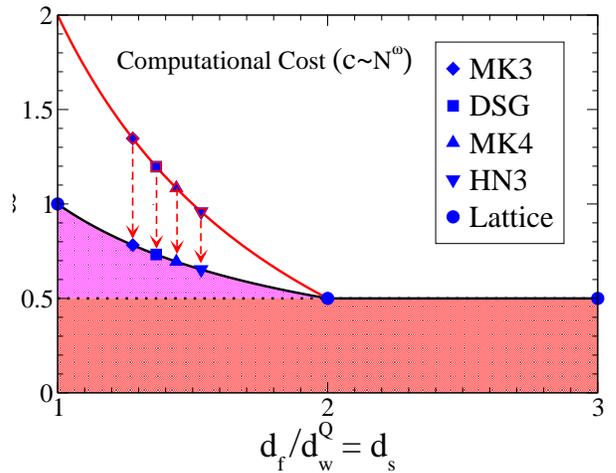}\hfill{}\caption{\label{fig:cplot}Illustration of the computational cost $c\sim N^{\omega}$
as a function of $d_{f}/d_{w}^{Q}$. The systems studied with RG all
possess $d_{f}/d_{w}^{Q}<2$, where the Grover limit ($\omega=\frac{1}{2}$)
can not be reached and the scaling is non-trivial. The naive Grover
search algorithm, analyzed in Sec.\ \ref{subsec:The-Quantum-search},
achieves the scaling in Eq.\ (\ref{eq:costNoTulsi}) (red line, red-framed
symbols), which can be optimized (down-arrows) by Tulsi's method~\cite{Tul08},
see Eq.\ (\ref{eq:CostTulsi}) (black line, blue symbols). Aside
from log-corrections, the RG finds $\omega=\max\left\{ \frac{d_{w}^{Q}}{d_{f}},\frac{1}{2}\right\} $,
which provides a fundamental limit, constraint by quantum transport
through the network geometry for $d_{f}/d_{w}^{Q}<2$ (magenta-shaded
area) or else by the inherent Grover limit of rotating the state vector
in Hilbert space~\cite{Bennett97} (red-shaded area). Assuming $d_{w}^{Q}=\frac{1}{2}d_{w}^{R}$,
as obtained in Ref.~\cite{Boettcher17a}, all results can be expressed
purely via the spectral dimension of the network Laplacian, for which
it is known that $d_{s}=2d_{f}/d_{w}^{R}$~\cite{Alexander82}. We
treat DSG as example here; the values for $d_{f}/d_{w}^{Q}$ listed
here for other networks -- MK3, MK4, and HN3, i.e., 3 and 4-regular
Migdal-Kadanoff and Hanoi networks -- are adapted from Tab.\ 1 in
Ref.~\cite{Boettcher14b}. Since $d_{w}^{Q}=1$ on a $d$-dimensional
lattice, i.e., $d_{s}=d_{f}=d$, this diagram applies directly to
lattices, with $d=2$ as the critical dimension~\cite{Pathria}.}
\end{figure}

Grover~\cite{Gro97a} has developed a quantum algorithm that, starting
from an initial state of uniform weight, can locate an entry in an
unordered database of $N$ elements with high probability in a time
that scales as $\sim\sqrt{N}$. This presents a quadratic speed-up
over classical search algorithms and has inspired countless algorithmic
developments \cite{ambainis_2004a,Aaronson05,Agliari2011,aharonov_1998a,Ambainis13,AKR05,chakraborty2015randomG,Childs04}
and recently several physical implementations \cite{Foulger14,Figgatt2017,Godfrin2017}.
In a database with a non-trivial network geometry, as in Fig.~\ref{fig:MarkedSite},
what we shall call a \emph{spatial} Grover search is faced with the
competition between
\begin{lyxlist}{00.00.0000}
\item [{(1)}] the accumulation of weight on a marked entry (or ``site'')
$w$ at the expense of its neighbors and 
\item [{(2)}] the ability to transport weight via quantum walk into that
neighborhood. 
\end{lyxlist}
Here we show how both of these tasks simultaneously can be described
(and optimized) with the real-space renormalization group (RG)~\cite{Pathria}.
As a result, see Fig.~\ref{fig:cplot}, we infer a lower bound on
the complexity (or asymptotic computational cost) of spatial Grover
search in terms of the network's fractal dimension $d_{f}$ and quantum
walk dimension $d_{w}^{Q}$ or, alternatively, it's spectral dimension
$d_{s}$. To this end, we study the exact RG on several fractal networks
exemplified by the dual Sierpinski gasket here; the corresponding
calculation for the other networks in Fig. \ref{fig:cplot} follows
from their RG in Refs. \cite{Boettcher14b,Boettcher17a}. Each of
these networks obtains the foregoing results in a non-trivial (and
often distinct \cite{Boettcher17a}) manner, which suggests (but does
not prove) that our prediction for the complexity bound exhibited
in Fig. \ref{fig:cplot} holds for networks of finite $d_{s}$ generally.
And although we assail fundamental tenets of computer science by exploring
the Grover algorithm where it \emph{fails} to saturate its optimal
limit, it is exactly in this regime, $1<d_{s}<2$, where we gain the
necessary insight to understand its behavior for all dimensions.

A discrete-time quantum walk with a coin was instrumental in the earliest
implementations of a quantum search algorithm to reach the Grover
limit ($\sim\sqrt{N}$) in as low as two dimensions~\cite{AKR05,PortugalBook},
up to logarithmic corrections, although alternative implementations
have been found~\cite{Aaronson05,Childs04b,Ambainis13}. While the
accumulation in (1) inherently~\cite{Bennett97} requires at least
$\sim\sqrt{N}$ updates, in (2) the neighborhood is replenished by
quantum transport on a time-scale of $\sim N^{d_{w}^{Q}/d_{f}}$,
as we will show. It becomes the limiting cost for the entire search
when $d_{f}<2d_{w}^{Q}$. The walk dimension $d_{w}(=d_{w}^{R})$
has been introduced for random walks as the exponent that characterizes
the asymptotic scaling relation between the spatial and temporal extend
in the probability density function~\cite{Havlin87,Redner01}, $\rho\left(x,t\right)\sim f\left(\left|x\right|^{d_{w}}/t\right)$.
Such a scaling is a powerful notion that in statistical physics has
lead to the invention of the Nobel prize winning idea of the renormalization
group (RG)~\cite{Kadanoff66,Wilson71}, as discussed in many textbooks~\cite{Pathria}.
We shall assume that such a scaling, now with some $d_{w}=d_{w}^{Q}$,
also exists for a the quantum walk with wave function $\psi_{x,t}$,
where $\rho\left(x,t\right)=\left|\psi_{x,t}^{2}\right|$. On a line,
so-called weak-limit results~\cite{konno_2003a} verify scaling with
$d_{w}^{Q}=1$, which has been reproduced with RG~\cite{Boettcher13a}.
This result, $d_{w}^{Q}=1$, has been extended to regular lattices
in all dimensions~\cite{grimmett_2004a}. The networks we consider
usually lack the translational invariance essential to prove properties
on lattice where $d_{f}=d$ is integer. Yet, our generalized results
for real (fractal) dimensions incorporate those for regular lattices.
They show that the Grover-limit can always be achieved in dimensions
$d>2$, where the average distance between sites on those lattices
is $\sim N^{d_{w}^{Q}/d}\ll\sqrt{N}$, and in the critical dimension
$d=2$ with likely logarithmic corrections. In turn, in the mean-field
limit~\cite{Pathria}, when all sites are neighbors (complete graph),
it is $d_{f}=\infty$ and transport is instantaneous, as it is for
random graphs of finite degree~\cite{chakraborty2015randomG} with
typical distances that are $\sim\ln N$.

The naive application of Grover's algorithm on a finite-dimensional
geometry also impacts the probability $p=\left|\psi_{w,t_{{\rm opt}}}^{2}\right|$
to overlap with the marked site $w$ -- the objective of the search
-- when the measurement is undertaken at the optimal time $t_{{\rm opt}}$.
The RG we discuss below finds asymptotically for large $N$ that $t_{{\rm opt}}\sim N^{d_{w}^{Q}/d_{f}}$,
accompanied by a decrease of $p\sim N^{1-2d_{w}^{Q}/d_{f}}$ when
$2d_{w}^{Q}/d_{f}>1$, which is comparable to the optimal overlap
with the target element found in a continuous-time quantum walk~\cite{Childs04,LiBo16}.
Thus, the complexity $c(N)$ of this naive quantum search algorithm,
which is given by the product of $t_{{\rm opt}}$ with the necessary
number of repeat-measurements ($\sim1/p$), becomes 
\begin{equation}
c=\frac{t_{{\rm opt}}}{p}\sim\max\left\{ \sqrt{N},N^{3\frac{d_{w}^{Q}}{d_{f}}-1}\right\} .\label{eq:costNoTulsi}
\end{equation}
We have verified the RG-predictions for both, $t_{{\rm opt}}$ and
$p$, for several other networks, see Fig.~\ref{fig:cplot}, and
with numerical simulations, explained in Fig.~\ref{fig:SearchCollapseDSG}.
Furthermore, an optimized algorithm was developed by Tulsi~\cite{Tul08}
that we can directly analyze with RG also. It allows to boost the
overlap $p$ at the expense of at most two extra qubits, when the
eigenvalue with the smallest positive argument of the evolution operator
fulfills certain properties. Then, the overlap always can be tuned
to a finite value, $p\sim1$, independent of $N$, and the complexity
bound finally attains its optimal form 
\begin{equation}
c_{{\rm Tulsi}}\sim\max\left\{ \sqrt{N},N^{\frac{d_{w}^{Q}}{d_{f}}}\right\} .\label{eq:CostTulsi}
\end{equation}
The dependence of the scaling of $c$ with $N$ on $d_{f}/d_{w}^{Q}$
for both of these scenarios is illustrated in Fig.~\ref{fig:cplot}.
Ultimately, our RG calculation below implies that the algorithmic
complexity is constrained by the speed of quantum transport: If $d_{f}/d_{w}^{Q}>2$,
Grover's limit can be reached! 

For coined quantum walks with no marked nodes it has been shown previously
that there is a relation between quantum walks and the corresponding
classical random walk~\cite{Boettcher14b,Boettcher17a}, i.e., $d_{w}^{Q}=\frac{1}{2}d_{w}^{R}$.
Using $d_{w}^{R}/d_{f}=2/d_{s}$~\cite{Alexander82}, we can represent
Eqs.~(\ref{eq:costNoTulsi}-\ref{eq:CostTulsi}) purely in spectral
terms, i.e., $d_{f}/d_{w}^{Q}=d_{s}$, as indicated in Fig.~\ref{fig:cplot}.
In that case, our result mirrors Szeged's finding for ``hitting times''
of $1/\sqrt{\delta}\sim N^{\frac{1}{d_{s}}}$ in bipartite networks
with spectral gap $\delta$ in quantized Markov chains\ \cite{Szegedy04,Krovi2015}.
A similar result has also been shown for quantum first passage times
\cite{Thiel18}. 

\begin{figure}
\vspace{-0.3cm}

\hfill{}\includegraphics[viewport=0bp 180bp 792bp 612bp,clip,width=1\columnwidth]{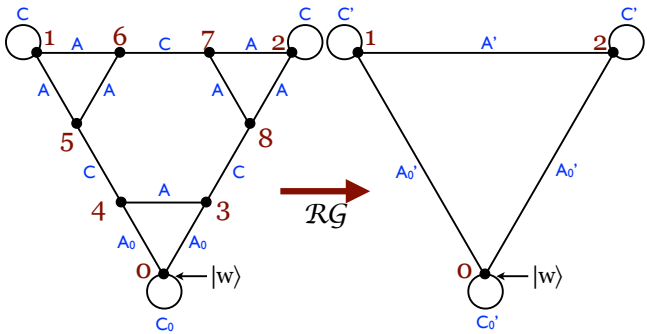}\hfill{}

\vspace{-0.3cm}
 \caption{\label{fig:DSGRGstep}Depiction of the (final) RG-step in the analysis
of DSG. Recursively, the inner-6 sites (here labeled $3,\ldots,8$)
of each larger triangle (left) in DSG are decimated to obtain a reduced
triangle (right) with renormalized hopping operators (primed). Since
site $w=0$ is distinct, modified recursion rules apply for the matrices
labeled with subscript $0$.}
\end{figure}

\section{Methods\label{sec:Methods}}

\subsubsection{Quantum Evolution Equation\label{sec:Quantum-Master}}

The time evolution of a quantum walk is governed by the discrete-time
equation 
\begin{equation}
\left|\Psi_{t+1}\right\rangle ={\cal U}\left|\Psi_{t}\right\rangle \label{eq:MasterEq}
\end{equation}
with unitary propagator ${\cal U}$. It resembles closely the master
equation for a random walk (or any other Markov process), for which
${\cal U}$ would be a stochastic operator. Then, in the discrete
$N$-dimensional site-basis $\left|x\right\rangle $ with $\psi_{x,t}=\left\langle x|\Psi_{t}\right\rangle $,
the probability density function is given by $\rho\left(x,t\right)=\left|\psi_{x,t}\right|^{2}$.
In this basis, the propagator can be represented as an $N\times N$
matrix ${\cal U}_{x,y}=\left\langle x\left|{\cal U}\right|y\right\rangle $
with operator-valued entries that describe the transitions between
neighboring sites (``hopping operators''). To study the long-time
dynamics, it is advantageous to apply a discrete Laplace transform\ \cite{Redner01},
\begin{equation}
\overline{\psi}_{x}\left(z\right)={\textstyle \sum_{t=0}^{\infty}}\psi_{x,t}z^{t},\label{eq:LaplaceT}
\end{equation}
such that Eq.~(\ref{eq:MasterEq}) becomes 
\begin{equation}
\overline{\psi}_{x}=\sum_{y}z{\cal U}_{x,y}\overline{\psi}_{y}+\psi_{x,t=0}.\label{eq:z_master}
\end{equation}

The self-similarity of fractal networks allows for a decomposition
of ${\cal U}_{x,y}$ into its smallest sub-structures, exemplified
by Fig.~\ref{fig:DSGRGstep}. It shows the elementary graph-let of
nine sites that is used to recursively construct the dual Sierpinski
gasket (DSG). The master equations pertaining to these sites are:
\begin{eqnarray}
\overline{\psi}_{0} & = & \left(M_{0}+C_{0}\right)\overline{\psi}_{0}+A\left(\overline{\psi}_{3}+\overline{\psi}_{4}\right)+I_{0}\psi_{IC},\label{eq:DSG_master}\\
\overline{\psi}_{\left\{ 1,2\right\} } & = & \left(M+C\right)\overline{\psi}_{\left\{ 1,2\right\} }+A\left(\overline{\psi}_{\left\{ 5,7\right\} }+\overline{\psi}_{\left\{ 6,8\right\} }\right)+I\psi_{IC},\nonumber \\
\overline{\psi}_{\left\{ 3,4\right\} } & = & M\overline{\psi}_{\left\{ 3,4\right\} }+C\overline{\psi}_{\left\{ 8,5\right\} }+A_{0}\overline{\psi}_{0}+A\overline{\psi}_{\left\{ 4,3\right\} }+I\psi_{IC},\nonumber \\
\overline{\psi}_{\left\{ 5,6,7,8\right\} } & = & M\overline{\psi}_{\left\{ 5,6,7,8\right\} }+C\overline{\psi}_{\left\{ 4,7,6,3\right\} }\nonumber \\
 &  & \qquad+A\left(\overline{\psi}_{\left\{ 1,5,2,7\right\} }+\overline{\psi}_{\left\{ 6,1,8,2\right\} }\right)+I\psi_{IC}.\nonumber 
\end{eqnarray}
The hopping operators $A$ and $C$ describe transitions between neighboring
sites, while $M$ (not shown in Fig.\ \ref{fig:DSGRGstep}) permits
the walker to remain on its site in a ``lazy'' walk. The inhomogeneous
$\psi_{IC}$-terms allow for an initial condition $\psi_{x,t=0}$
on the respective site $x$.

Preserving the norm of the quantum walk demands unitary propagation,
i.e., $\mathbb{I}={\cal U}^{\dagger}{\cal U}$. This can be achieved
in the discrete-time case only when the hopping operators like $\left\{ A,C,M\right\} $
in Eqs.\ (\ref{eq:DSG_master}) are matrices, not scalars. Correspondingly,
the state of the walk at each site, $\psi_{x,t}$, must be a vector
of conforming length. Each update, a conforming coin matrix ${\cal C}$
entangles the components of the state vector, which the hopping operators
subsequently distribute to their respective neighboring site. For
coined quantum walks, it has been conventional to consider merely
those coins whose dimensions adhere to the degree of the sites in
the network under investigation. Then, each component of a site's
state vector is shifted along one specific direction at each update,
ensuring the unitarity of the propagator ${\cal U}$ overall. However,
for networks of higher degree, or of mixed degree, this approach becomes
quite unwieldy, if not impossible. In Appendix A, we have laid out
how to obtain generalized unitarity conditions for any network. When
applied to DSG specifically, we have derived the following conditions
concerning the hopping operators in Eqs.~(\ref{eq:DSG_master}):
\begin{eqnarray}
\mathbb{I} & = & A^{\dagger}A+B^{\dagger}B+C^{\dagger}C+M^{\dagger}M,\label{eq:UnitarityDSG}\\
0 & = & A^{\dagger}B+B^{\dagger}M+M^{\dagger}A=C^{\dagger}M+M^{\dagger}C=A^{\dagger}C=B^{\dagger}C.\nonumber 
\end{eqnarray}
These conditions at hand, we can now systematically design generalized
hopping operators $\left\{ A,B,C,M\right\} $. We make a most simple
choice by requiring an additional symmetry, $A=B$, while choosing
$2\times2$-matrices 
\begin{equation}
M=\left[\begin{array}{cc}
-\frac{1}{3} & 0\\
0 & 0
\end{array}\right]{\cal C},\quad A=\left[\begin{array}{cc}
\frac{2}{3} & 0\\
0 & 0
\end{array}\right]{\cal C},\quad C=\left[\begin{array}{cc}
0 & 0\\
0 & 1
\end{array}\right]{\cal C},\label{eq:MAC_IC}
\end{equation}
that satisfy Eqs.\ (\ref{eq:UnitarityDSG}) for any unitary coin
${\cal C}$. Here, the most general unitary $2\times2$ coin matrix
is given by 
\begin{equation}
{\cal C}=\left(\begin{array}{cc}
\sin\eta & e^{i\chi}\,\cos\eta\\
e^{i\vartheta}\,\cos\eta & -e^{i\left(\chi+\vartheta\right)}\,\sin\eta
\end{array}\right).\label{eq:HadamardCoin}
\end{equation}

In the following, we merely consider variable $\eta$ but set $\chi=\vartheta=0$.
{[}We note that for non-zero $\chi$ and $\vartheta$, the following
results would be identical aside from a trivial rotation in the Laplace
parameter, $z\to z\,e^{-i\left(\chi+\vartheta\right)}$.{]} However,
with the free parameter $\eta$, which specifies the extend by which
the components of the state vector get entangled, we are now in a
position to study an entire family of problems. Even though the degree
of the network is larger than this coin-space, for the Hadamard coin
in Eq.\ (\ref{eq:HadamardCoin}) we show in the following that it
reproduces the phenomenology of the quantum walk with $3\times3$-matrices
and lower symmetry ($A\not=B$) for the Grover coin described in Refs.\ \cite{QWNComms13,Boettcher17a}.
Besides this ``minimalist'' example, other interesting $3$ (or
higher) dimensional matrices that solve the conditions in Eq.\ (\ref{eq:UnitarityDSG})
may exist, potentially harboring new universality classes and localization
behaviors\ \cite{Falkner14a}.

In Eqs.~(\ref{eq:DSG_master}), we have distinguished site $w=0$.
(This choice is largely a matter of convenience; any other $w$ would
result in the same scaling but with a $w$-dependent pre-factor \cite{SWlong}.)
In such a way, we can study either a quantum walk starting on that
site to determine the spreading dynamics or the quantum search problem
of amplifying the wave-function on site $w=0$ after starting from
a uniform initial state $\psi_{x,0}=\frac{1}{\sqrt{N}}\psi_{IC}$,
where $\psi_{IC}$ denotes an initial state-vector. The latter case
is discussed below. In the former case, the initial condition is localized
at $w$, $\psi_{x,t=0}=\delta_{x,w}\psi_{IC}$, with $I_{0}=\mathbb{I}$,
$I\equiv0$, $M_{0}=M$, $A_{0}=A$, and $C_{0}=C$, as discussed
elsewhere\ \cite{Boettcher17a}. Although they result in very different
physical situations, both cases built on the following analysis of
the RG-recursions for the homogeneous walk, irrespective of the initial
conditions $I$.

\subsubsection{RG for the Homogeneous Quantum Walk \label{sec:The-Quantum-Walk}}

As we have indicated in the introduction, the real-space RG for a
walk\ \cite{Redner01} provides information that relates the temporal
and spatial spreading of the walk. Instead of yielding a specific,
quantitative result on a question of, say, ``How much time $T$,
on average, does it take for a walk to fall off a table of base-length
$L$ after starting in its center?'', the RG answers the \emph{scaling}
question ``By how much does a change in $L_{k}\to L_{k+1}=2L_{k}$
rescale time $T_{k}\to T_{k+1}=\lambda T_{k}$?'' in each step $k\to k+1$
of the RG. Assuming scaling $T_{k}\sim L_{k}^{d_{w}}$ (at least asymptotically
for all large $k$), the answer to that question would imply $d_{w}=\log_{2}\lambda$.
Clearly, for a classical random walk (i.e., diffusion) on any $d$-dimensional
``table'' it is $\lambda=4$, i.e., $d_{w}=2$. In a fractal geometry,
the answer to this question generally is non-trivial\ \cite{Havlin87,Redner01}.
This example illustrates the relevance of RG for the complexity of
the Grover algorithm which concerns the question on ``How much does
$t_{{\rm opt}}$ for search increase when I increase $N(=L^{d_{f}})$''.
Note, however, that due to the Laplace transform in Eq.\ (\ref{eq:LaplaceT})
the large-$t$ limit is accessed for $z\to1$ here.

The recursive structure of DSG (and many other fractals, such as those
discussed in Ref.\ \cite{Boettcher14b}), allows to establish exact
recursion relations between a walk at length $k$ and $k+1$. These
RG-recursions for the DSG, as represented by Eqs.~(\ref{eq:RGrecur}),
are generic and have been derived previously\ \cite{Boettcher17a}.
In Appendix B, we recall how to obtain those recursions, for completeness.
Iterating these RG-recursions as described there for only one step
already reveals a recursive pattern that suggests the parametrization
\begin{align}
M_{k} & =\left(a_{k}-\frac{2}{3}b_{k}\right)\left[\begin{array}{cc}
1 & 0\\
0 & 0
\end{array}\right]{\cal C},\nonumber \\
A_{k} & =\left(a_{k}+\frac{1}{3}b_{k}\right)\left[\begin{array}{cc}
1 & 0\\
0 & 0
\end{array}\right]{\cal C},\label{eq:MAC_k}\\
C_{k} & =\,z\,\left[\begin{array}{cc}
0 & 0\\
0 & 1
\end{array}\right]{\cal C},\nonumber 
\end{align}
which \emph{exactly} closes on itself after one iteration, $k\to k+1$,
when we identify for the \emph{scalar} RG-flow: 
\begin{eqnarray}
a_{k+1} & = & \frac{\begin{array}{l}
\left(9a_{k}b_{k}+3z\,a_{k}-2z\,b_{k}\right)\sin\eta\\
\quad+9z\,a_{k}b_{k}+3a_{k}-2b_{k}
\end{array}}{\begin{array}{l}
3\left(3z-6a_{k}+b_{k}\right)\sin\eta\\
\quad+3(3-6z\,a_{k}+z\,b_{k})
\end{array}},\label{eq:RGflowDSG}\\
\nonumber \\
b_{k+1} & = & \frac{\begin{array}{l}
2\left(9a_{k}b_{k}^{2}+3z\,a_{k}b_{k}+z\,b_{k}^{2}-3z^{2}a_{k}+2z^{2}b_{k}\right)\sin^{2}\eta\\
\quad+4\left(1-z^{2}\right)\left(6a_{k}b_{k}-b_{k}^{2}\right)\sin\eta\\
\quad+2\left(3a_{k}-2b_{k}-3z\,a_{k}b_{k}-z\,b_{k}^{2}-9z^{2}a_{k}b_{k}^{2}\right)
\end{array}}{\begin{array}{l}
2\left(6a_{k}b_{k}-b_{k}^{2}+3z\,a_{k}+z\,b_{k}+3z^{2}\right)\sin^{2}\eta\\
\quad+4\left(1-z^{2}\right)\left(3a_{k}-2b_{k}\right)\sin\eta\\
\quad-2\left(3+3z\,a_{k}+z\,b_{k}+6z^{2}a_{k}b_{k}-z^{2}b_{k}^{2}\right)
\end{array}}.\nonumber 
\end{eqnarray}
This flow is initiated at $k=0$ with $a_{k=0}=z/3$ and $b_{k=0}=z$,
to match Eqs.~(\ref{eq:MAC_k}) to the unrenormalized hopping operators
in Eqs.~(\ref{eq:MAC_IC}). Note that these RG-flow recursions are
vastly simpler than the 5-term recursions previously reported in Ref.~\cite{QWNComms13},
or those in Ref.~\cite{Boettcher17a}, even though here they describe
an entire family of coins via the coin-parameter $\eta$.

As explained above, the real-space RG equations encapsulate the behavior
of the physical process under rescaling of length (on DSG, from base-length
$L_{k}=2^{k}$ to $L_{k+1}=2L_{k}$, while size $N_{k}=L_{k}^{d_{f}}=3^{k}$
changes by a factor of 3, i.e., $d_{f}=\log_{2}3$). Thus, we now
proceed to study the fixed-point properties of the RG-flow in Eq.\ (\ref{eq:RGflowDSG})
at $k\sim k+1\to\infty$ near $z\to1$~\cite{Redner01}. The particular
combination of $a_{k}$ and $b_{k}$ in Eq.~(\ref{eq:MAC_k}) ensures
that the Jacobian of the fixed point already is diagonal, with eigenvalues
$\lambda_{1}=3$ and $\lambda_{2}=\frac{5}{3}$. Extending the expansion
of Eq.~(\ref{eq:RGflowDSG}) in powers of $\zeta=z-1$ for $k\to\infty$
to sufficiently-high order, we obtain: 
\begin{eqnarray}
a_{k}\left(z\right) & \sim & \frac{1}{3}+\zeta\thinspace{\cal A}\lambda_{1}^{k}+\zeta^{2}\alpha_{k}^{(2)}+\zeta^{3}\alpha_{k}^{(3)}+\ldots,\nonumber \\
b_{k}\left(z\right) & \sim & 1+\zeta\thinspace{\cal B}\lambda_{2}^{k}+\ldots,\label{eq:abFP}
\end{eqnarray}
with unknown constants ${\cal A}$ and ${\cal B}$. Here, we defined
\begin{eqnarray}
\alpha_{k}^{(2)} & \sim & \frac{3}{2}{\cal A}^{2}\lambda_{1}^{2k}+\ldots,\label{eq:alphak}\\
\alpha_{k}^{(3)} & \sim & \frac{9}{4}{\cal A}^{3}\lambda_{1}^{3k}-\frac{3}{8}{\cal A}^{2}{\cal B}\lambda_{1}^{2k}\lambda_{2}^{k}+\ldots,\nonumber 
\end{eqnarray}
where we have only kept leading-order terms relevant for the following
considerations. It was argued previously~\cite{Boettcher16,Boettcher17a}
that we can identify: 
\begin{equation}
d_{f}=\log_{2}\lambda_{1},\qquad d_{w}^{Q}=\log_{2}\sqrt{\lambda_{1}\lambda_{2}},\label{eq:dfw}
\end{equation}
i.e., $d_{f}=\log_{2}3$ and $d_{w}^{Q}=\log_{2}\sqrt{5}$ for DSG.

\section{Results for the Complexity of Quantum Search\label{subsec:The-Quantum-search}}

To apply the RG results in Sec.~\ref{sec:Methods} to the corresponding
quantum search problem, we use the \emph{abstract search algorithm}\ \cite{Gro97a,AKR05,PortugalBook}.
It replaces the operator ${\cal U}$ by an equally unitary ``search''-propagator
${\cal U}_{w}={\cal U}\cdot{\cal R}_{w}$ that distinguishes the sought-after
site $\left|w\right\rangle $ from the remaining sites using the search-operator
\begin{equation}
{\cal R}_{w}=\mathbb{I}-\left|w\right\rangle \left\langle w\right|\left(2D\right).\label{eq:Rw}
\end{equation}
The walk operator ${\cal U}$ corresponds to the inversion-about-average
operator defined by Grover\ \cite{Gro97a}. It ``drives'' the quantum
walk by transporting the weight of the wave-function between neighboring
sites in an attempt to make it uniform. Alas, in the quantum search,
which starts from a uniform state, the prior reflection of the phase
at site $w$ by ${\cal R}_{w}$ first imbalances the amplitude there,
before ${\cal U}$ now amplifies this imbalance at $w$. Thus, site
$w$ acts as an ``attractor'' for the weight of the wave-function
at the expense of its immediate neighbors - a deficit that ${\cal U}$
persistently tries to correct. Since we require ${\cal U}_{w}$ to
be unitary, so must be ${\cal R}_{w}$ in Eq.\ (\ref{eq:Rw}), which
implies the condition
\begin{equation}
2D^{\dagger}D=D^{\dagger}+D.\label{eq:Dcondition}
\end{equation}
Grover\ \cite{Gro97a}, and by default many authors since, have further
imposed reflectivity, ${\cal R}_{w}^{2}=\mathbb{I}$, which conveniently
reduces Eq.\ (\ref{eq:Dcondition}) to $D=D^{2}$, further implying
hermiticity, $D=D^{\dagger}$. These conditions on $D$ still allow
for entire classes of operators, as well as $D=\mathbb{I}$. We will
consider first the family, 
\begin{equation}
D(\gamma)=\left[\begin{array}{cc}
\cos^{2}\gamma & \sin\gamma\cos\gamma\\
\sin\gamma\cos\gamma & \sin^{2}\gamma
\end{array}\right],\label{eq:Dmatrix}
\end{equation}
which for $\gamma=\frac{\pi}{4}$ reduces to the Grover operator that
is widely used in numerical simulations for this task\ \cite{PortugalBook}.
Note that $D$ in Eq.\ (\ref{eq:Dmatrix}) is singular, $\det D=0$,
for all $\gamma$, while $D=\mathbb{I}$ is the unique non-singular
solution of $D=D^{2}$. The RG reveals that $D=\mathbb{I}$ does not
allow for an efficient search, as we will show in Sec.\ \ref{subsec:Discussion-of-D=00003DI}.
Similarly, the RG calculation in Sec.\ \ref{subsec:Discussion-of-Non-Reflective}
implies that reflectivity appears to be necessary condition.

\begin{figure*}
\hfill{}%
\begin{minipage}[b][1\totalheight][s]{0.3\textwidth}%
\includegraphics[viewport=50bp 0bp 600bp 340bp,clip,angle=270,origin=lb,width=0.8\textwidth]{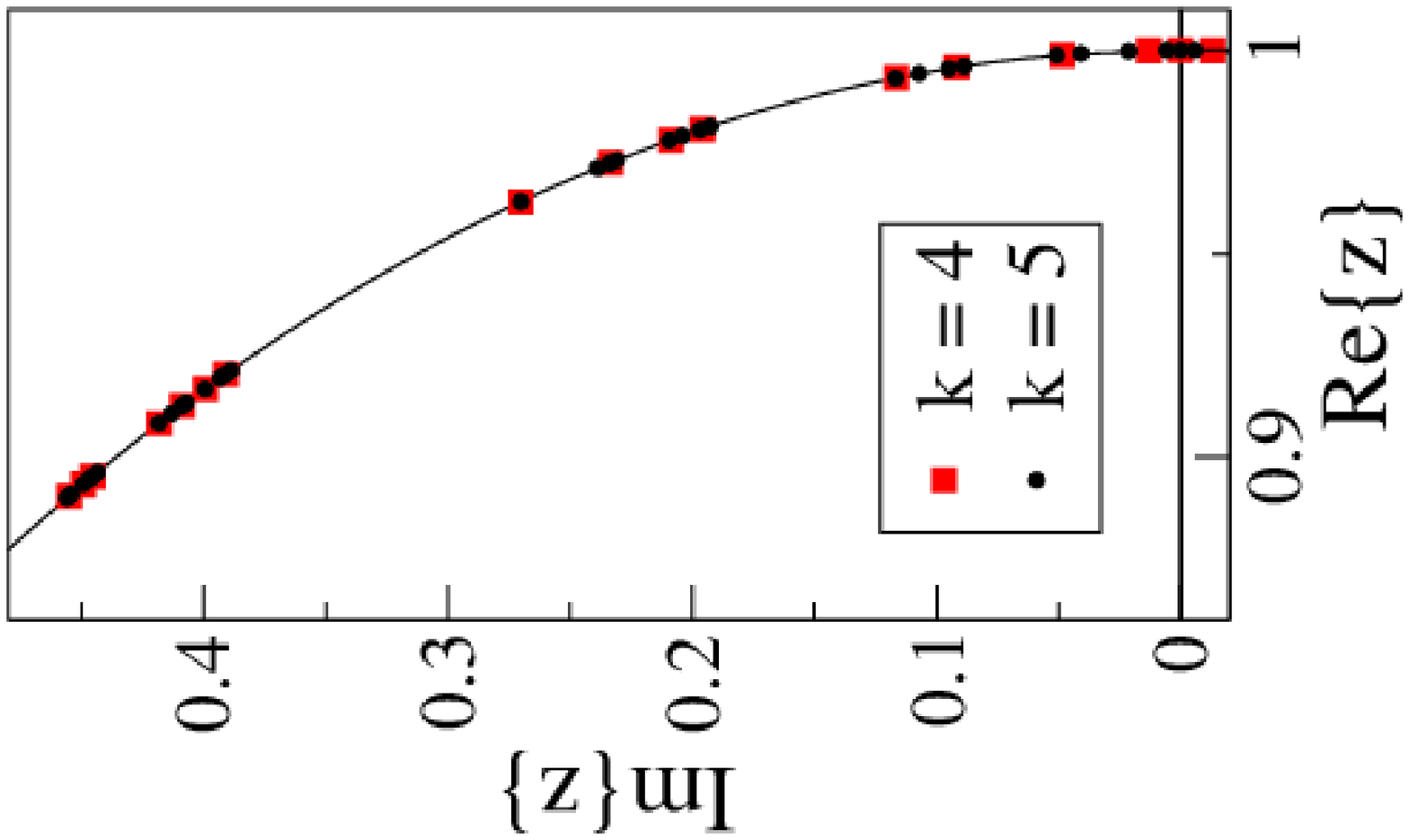}\hfill{}\caption{\label{fig:DSGpoles}Plot of the poles of the Laplace transform for
the amplitude at the sought-for site, $\overline{\psi}_{0}^{(k)}(z)$
in Eq.~(\ref{eq:DSGPsi0}), in the complex-$z$ plane at RG-steps
$k=4$ ($\blacksquare$) and $k=5$ ($\CIRCLE$) for quantum search
on the dual Sierpinski gasket (DSG). (The poles are certain to occur
in complex-conjugate pairs, so only the upper $z$-plane is shown.)
A finite fraction of those poles progressively impinge on the real-$z$
axis at $z=1$. }
\end{minipage}\hfill{}%
\begin{minipage}[b][1\totalheight][s]{0.65\textwidth}%
\hfill{}\includegraphics[viewport=0bp 10bp 725bp 550bp,clip,width=0.45\textwidth]{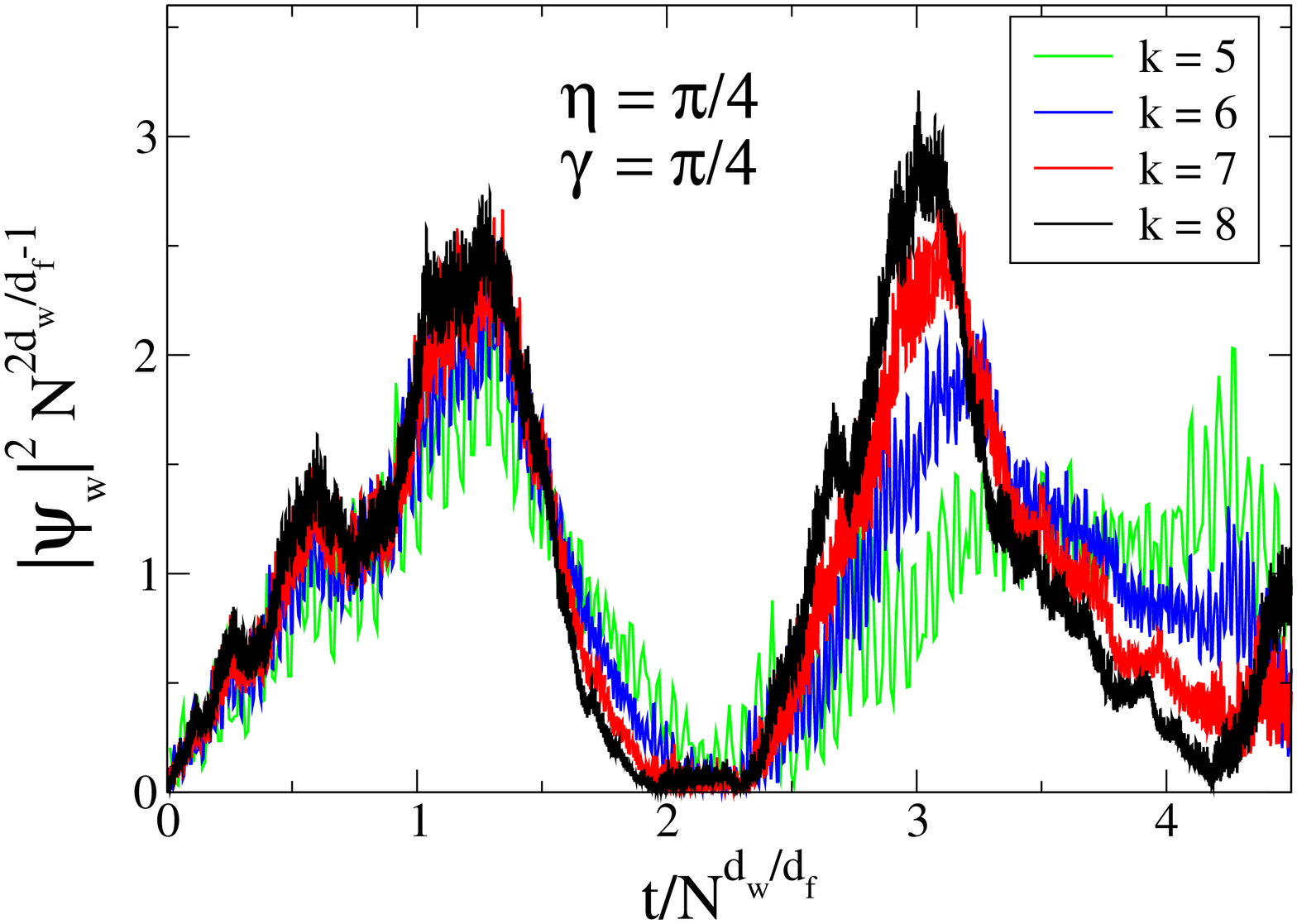}\hfill{}\includegraphics[viewport=0bp 10bp 725bp 550bp,clip,width=0.45\textwidth]{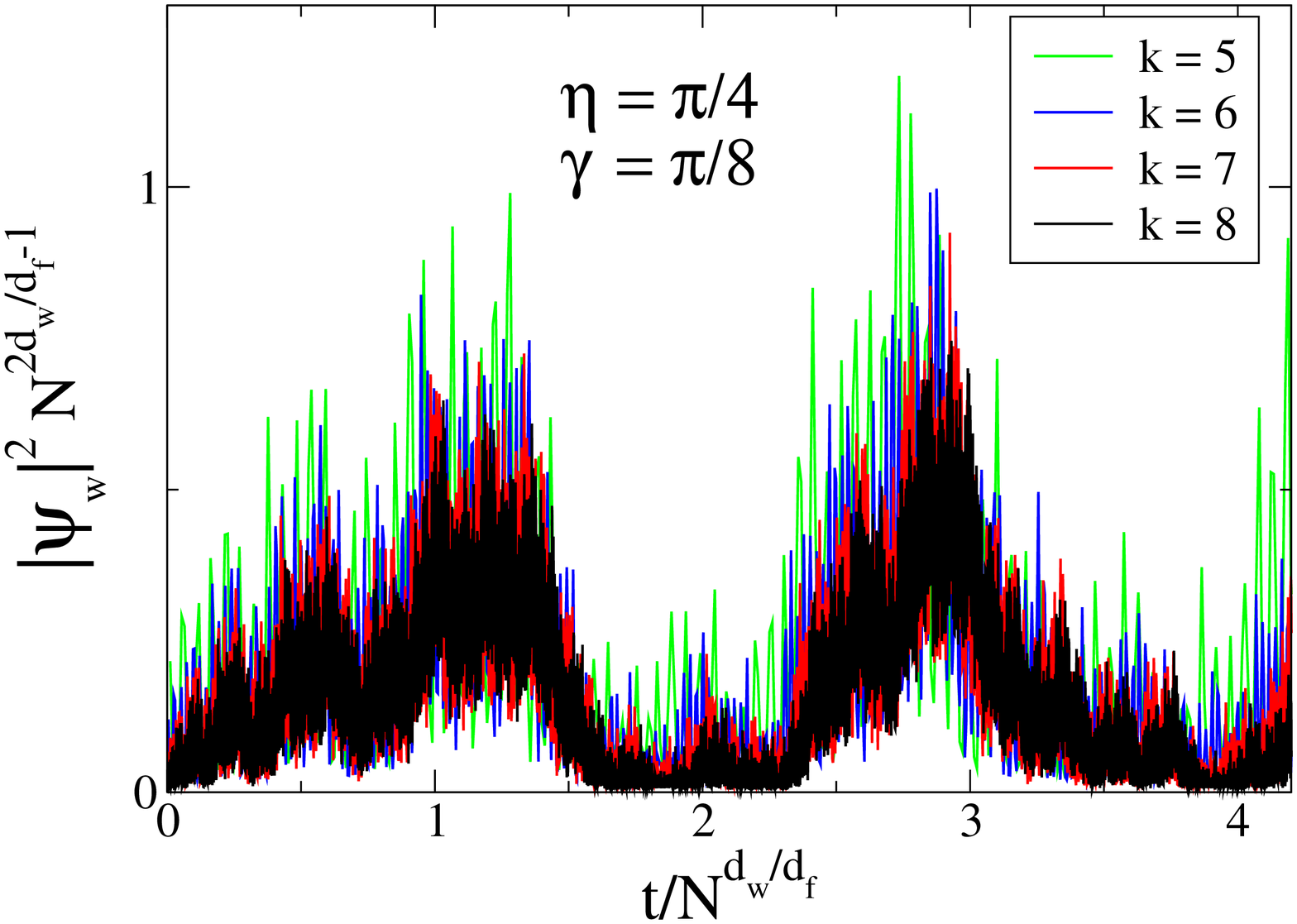}\hfill{}

\hfill{}\includegraphics[viewport=0bp 10bp 725bp 550bp,clip,width=0.45\textwidth]{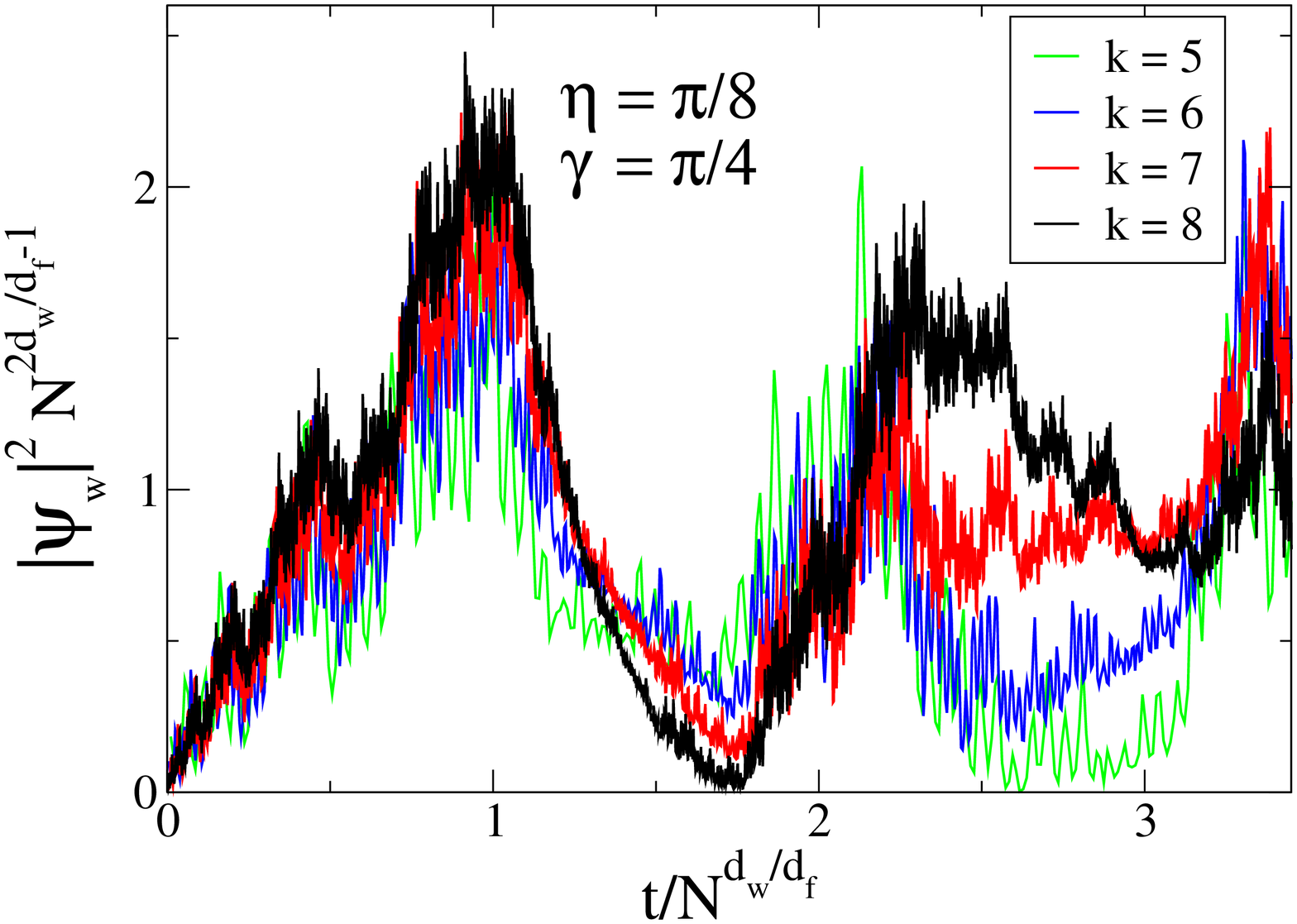}\hfill{}\includegraphics[viewport=0bp 10bp 725bp 550bp,clip,width=0.45\textwidth]{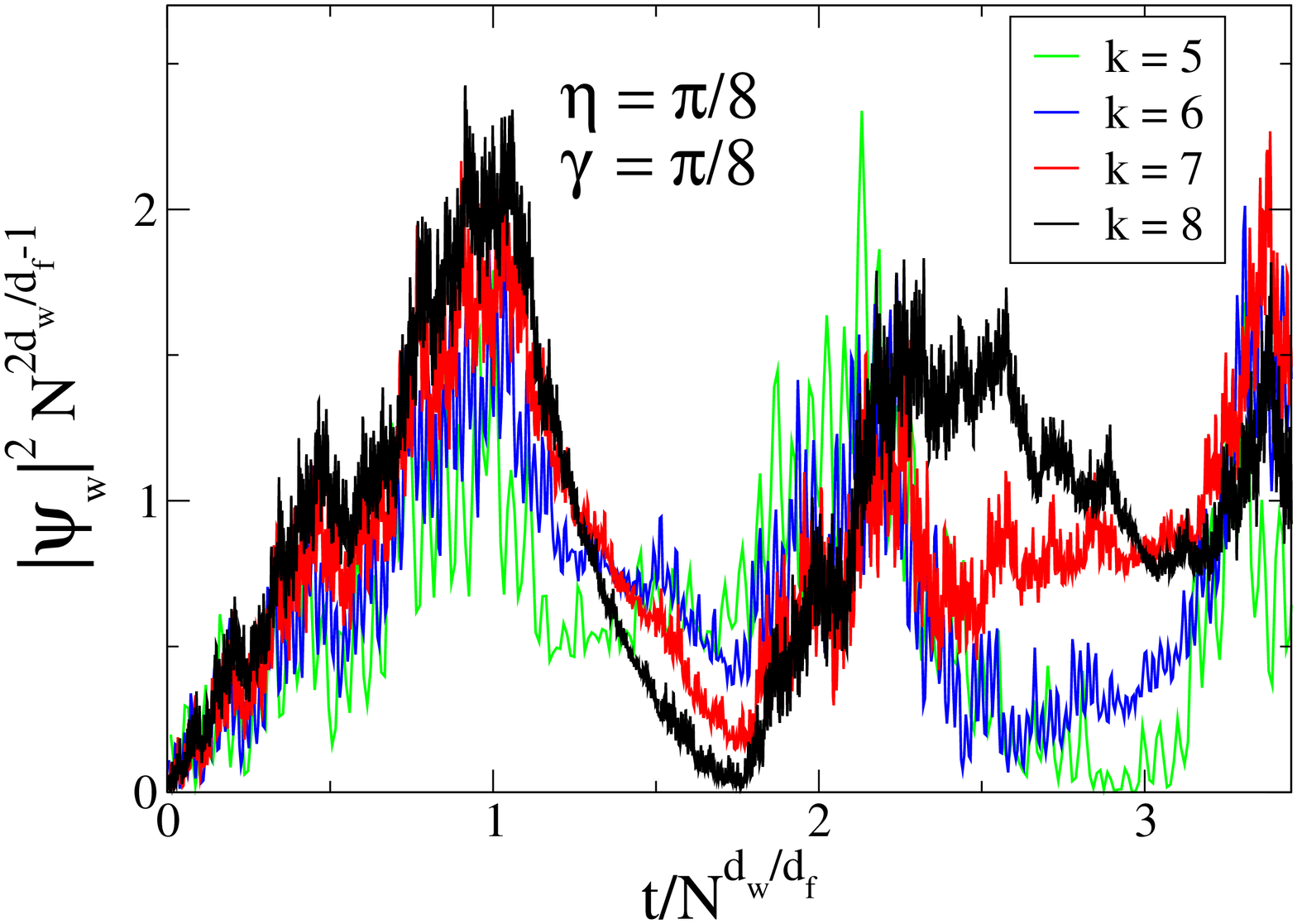}\hfill{}

\caption{\label{fig:SearchCollapseDSG}Plot of the probability $p=\left|\psi_{w,t}^{2}\right|$
to detect the quantum walk at site $w=0$ as a function of time $t$
for all combinations of parameters of the coin in Eq.\ (\ref{eq:HadamardCoin})
with $\eta=\pi/4$ and $\pi/8$ and of the search operator in Eq.\ (\ref{eq:Dmatrix})
with $\gamma=\pi/4$ and $\pi/8$ for DSG of size $N_{k}=3^{k}$.
(Shown here are $k=5,\ldots,8$, in order from bottom to top near
the first peak at $t/N^{d_{w}^{Q}/d_{f}}\approx1$.) Appropriately
rescaled according to Eq.~(\ref{eq:mainresult}), the data collapses
on a quasi-periodic sinusoidal function, as in Eq.~(\ref{eq:psi0t}).
For the collapse, we use the value for $d_{w}^{Q}$ and $d_{f}$ according
to Eq.~(\ref{eq:dfw}). The optimal time for a measurement would
be at $t_{{\rm opt}}/N^{d_{w}^{Q}/d_{f}}\approx1$, near the first
well-formed peak. Smoother (but less insightful) behavior for $p$
would be obtained after implementing Tulsi's method\ \cite{Tul08}
(see also Sec.\ \ref{subsec:Extension-to-Tulsi's} and the Appendix
D) and various improvements of the evolution operator\ \cite{Patel12,Marquezino12}.}
\end{minipage}
\end{figure*}

\subsection{General Considerations for Quantum Search on DSG\label{subsec:General-Considerations-for-DSG}}

Uniform initial conditions are provided by $\left|\Psi_{t=0}\right\rangle =\frac{1}{\sqrt{N}}\sum_{x}\left|x\right\rangle \otimes\left|\psi_{IC}\right\rangle $,
i.e., $\psi_{x,t=0}=\frac{1}{\sqrt{N}}\psi_{IC}$. With the goal to
optimize the amplitude $\psi_{0,t}$ to detect the walk on the sought-after
site $w=0$ in the shortest time possible, Eq.~(\ref{eq:z_master})
then becomes 
\begin{align}
\overline{\psi}_{x} & =\sum_{y}z\left({\cal U}_{w}\right)_{x,y}\overline{\psi}_{y}+\frac{\psi_{IC}}{\sqrt{N}},\label{eq:SearchU}\\
 & =\sum_{y}z{\cal U}_{x,y}\left(\mathbb{I}-2D\delta_{y,0}\right)\overline{\psi}_{y}+\frac{\psi_{IC}}{\sqrt{N}},\nonumber 
\end{align}
which turns into Eqs.~(\ref{eq:DSG_master}) when applied to the
DSG with ${\cal O}_{0}={\cal O}\left(\mathbb{I}-2D\right)$ for each
${\cal O}\in\left\{ A,C,M,I\right\} $. After $k$ iterations, in
the final step, as shown in Fig.~\ref{fig:DSGRGstep}, the DSG reduces
to a triangle of sites with: 
\begin{eqnarray}
\overline{\psi}_{0} & = & \left(M_{k}+C_{k}\right)\left(\mathbb{I}-2D\right)\overline{\psi}_{0}+A_{k}\left(\overline{\psi}_{1}+\overline{\psi}_{2}\right)\nonumber \\
 &  & \quad+I_{k}\left(\mathbb{I}-2D\right)\frac{\psi_{IC}}{\sqrt{N}},\label{eq:psiDSGtriangle}\\
\overline{\psi}_{\left\{ 1,2\right\} } & = & \left(M_{k}+C_{k}\right)\overline{\psi}_{\left\{ 1,2\right\} }+A_{k}\left[\left(\mathbb{I}-2D\right)\overline{\psi}_{0}+\overline{\psi}_{\left\{ 2,1\right\} }\right]\nonumber \\
 &  & \quad+I_{k}\frac{\psi_{IC}}{\sqrt{N}}.\nonumber 
\end{eqnarray}
Solving for $\overline{\psi}_{0}$, we obtain 
\begin{equation}
\overline{\psi}_{0}=\left[\mathbb{I}-\left(M_{k}+C_{k}+V_{k}A_{k}\right)\left(\mathbb{I}-2D\right)\right]^{-1}\left(\mathbb{I}+V_{k}\right)I_{k}\frac{\psi_{IC}}{\sqrt{N}}\label{eq:psi0simple}
\end{equation}
where we abbreviated $V_{k}=2A_{k}\left(\mathbb{I}-A_{k}-C_{k}-M_{k}\right)^{-1}$.

Note that $\overline{\psi}_{0}$ appears to depend also on the RG-recursion
for $I_{k}$. Yet, we can eliminate it by the following consideration:
If it were $D=0$, then we would have ${\cal R}_{w}=\mathbb{I}$ and
${\cal U}_{w}\equiv{\cal U}$ for the propagator, which would leave
the uniform initial state \emph{invariant}. Thus, $\left.\overline{\psi}_{0}\left(z\right)\right|_{D=0}=F\left(z\right)\frac{\psi_{IC}}{\sqrt{N}}$
, where $F(z)$ has at most $N$-independent, trivial poles. In fact,
we find from Eq.~(\ref{eq:psi0simple}) at $D=0$ that
\begin{eqnarray}
F\left(z\right) & = & \left[\mathbb{I}-\left(M_{k}+C_{k}+V_{k}A_{k}\right)\right]^{-1}\left(\mathbb{I}+V_{k}\right)I_{k},\nonumber \\
 & = & \frac{1}{1-z^{2}}\left[\begin{array}{cc}
1+z\,\sin\eta & z\,\cos\eta\\
z\,\cos\eta & 1-z\,\sin\eta
\end{array}\right],\label{eq:Fz}
\end{eqnarray}
independent of $k$. Then substituting Eq.~(\ref{eq:Fz}) back into
Eq.\ (\ref{eq:psi0simple}) yields 
\begin{equation}
\overline{\psi}_{0}=\left[\mathbb{I}-2G\left(z\right)D\right]^{-1}F\left(z\right)\frac{\psi_{IC}}{\sqrt{N}},\label{eq:DSGPsi0}
\end{equation}
with 
\begin{equation}
G\left(z\right)=\left[\mathbb{I}-\left(M_{k}+C_{k}+V_{k}A_{k}\right)^{-1}\right]^{-1}.\label{eq:Gz}
\end{equation}

Even before we discuss the effect of the search-operator $D$, the
properties of $G(z)$ itself are crucial for the proper interpretation
of the quantum search. It closely resembles the Laplace-space amplitude
for a quantum walker to remain at its starting location examined previously\ \cite{Boettcher17a},
although that situation has quite different (localized) initial conditions.
Inserting the RG-results from Eqs.~(\ref{eq:MAC_k}-\ref{eq:alphak})
into Eq.~(\ref{eq:Gz}), we find in powers of $\zeta=z-1$: 
\begin{eqnarray}
G(z) & \sim & \frac{1}{9{\cal A}\lambda_{1}^{k}}G_{-1}\zeta^{-1}+G_{0}\zeta^{0}+\frac{5{\cal B}\lambda_{2}^{k}}{24}G_{1}\zeta^{1}+\ldots.\label{eq:Gzetaseries}
\end{eqnarray}
with dominant contributions in large-$k$ from the matrices 
\begin{eqnarray}
G_{-1} & = & \left[\begin{array}{cc}
1 & \frac{\cos\eta}{1+\sin\eta}\\
\frac{\cos\eta}{1+\sin\eta} & \frac{1-\sin\eta}{1+\sin\eta}
\end{array}\right],\label{eq:G_i}\\
G_{0} & = & \frac{1}{2}\left[\begin{array}{cc}
1 & \frac{\cos\eta}{1+\sin\eta}\\
-\frac{\cos\eta}{1+\sin\eta} & 1
\end{array}\right]+\ldots,\nonumber \\
G_{1} & = & G_{-1}+\ldots.\nonumber 
\end{eqnarray}
The emergence of $\lambda_{2}^{k}$ as the dominant term for large
$k$ at order $\zeta^{1}$ is a consequence of unitarity\ \cite{Boettcher17a},
due to a delicate cancellation between $\alpha_{2}^{2}$ and $\left({\cal A}\lambda_{1}^{k}\right)\alpha_{3}$
in Eq.~(\ref{eq:alphak}). Also, in the following it will prove crucial
that $G_{-1}$ in Eq.\ (\ref{eq:G_i}) is a \emph{singular} matrix. 

\subsection{Discussion of $D$ in Eq.\ (\ref{eq:Dmatrix})\label{subsec:Discussion-of-GroverD}}

With a search operator containing the generalized Groverian matrix
$D(\gamma)$ in Eq.\ (\ref{eq:Dmatrix}), we indeed find a quantum
search algorithm with a non-trivial complexity. With $G(z)$ in Eq.\ (\ref{eq:Gzetaseries}),
we can construct the combination $\mathbb{I}-2G\left(z\right)D$ in
Eq.~(\ref{eq:DSGPsi0}), which itself is singular at order $\zeta^{-1}$,
due to $G_{-1}$. It is thus not surprising to find that its inverse
in Eq.~(\ref{eq:DSGPsi0}) has a leading contribution of order $\zeta^{0}$.
The combination of $\left[\mathbb{I}-2G\left(z\right)D\right]^{-1}F\left(z\right)$
in Eq.~(\ref{eq:DSGPsi0}) should therefore be $\sim\zeta^{-1}$,
owing to the pole in $F(z)$. Amazingly, however, the matrix $F(z)$
in Eq.~(\ref{eq:Fz}) exactly \emph{annihilates} that $\zeta^{-1}$-term
in $\overline{\psi}_{0}$ for any $\eta$ or $\gamma$. Evaluation
of Eq.~(\ref{eq:DSGPsi0}) then leads to: 
\begin{equation}
\overline{\psi}_{0}\sim\left\{ \left({\cal A}\lambda_{1}^{k}\right)F_{0}\zeta^{0}+\left({\cal A}\lambda_{1}^{k}\right)^{2}\left({\cal B}\lambda_{2}^{k}\right)F_{2}\zeta^{2}+\ldots\right\} \frac{\psi_{IC}}{\sqrt{N}}\label{eq:psi0zeta}
\end{equation}
where we have only kept the most-divergent term in $k$ at each order
of $\zeta$. Each term contains a $k$-independent matrix $F_{i}\left(\eta,\gamma,\zeta\right)$
that is regular in $\zeta$ and that captures the entire dependence
on the coin-parameter $\eta$ from Eq.~(\ref{eq:HadamardCoin}) and
the $\gamma$-dependence of the search operator $D$ in Eq.~(\ref{eq:Dmatrix}).
Although each such matrix is singular, every one of their components
is a well-behaved function on $0<\eta<\frac{\pi}{2}$ and $0<\gamma<\frac{\pi}{2}$
without poles or selections for which any $F_{i}$ would vanish entirely.
Thus, we can conclude that our following results for the scaling of
quantum search are \emph{universal} as far as this choice of coin
and search operator is concerned. 

To extract the relevant scaling behavior for the amplitude at the
sought-for site, $\overline{\psi}_{0}$ in Eq.~(\ref{eq:psi0zeta}),
we have to discuss the expectation we have for its form\ \cite{Redner01}.
For $t>0$, $\psi_{0,t}$ should be a periodic function of some fundamental
period $T(N)=2\pi/\theta_{k}$ that is small at $t=0$ but rises to
a significant maximum with some amplitude-factor $\sim N^{\epsilon}$
at the optimal time to conduct a measurement, $t_{{\rm opt}}=T(N)/4$.
Both, the increasing number of Laplace-poles of the RG with increasing
system size, shown in Fig.\ \ref{fig:DSGpoles}, and the additional
``overtones'' exhibited in the numerical simulations in Fig.\ \ref{fig:SearchCollapseDSG},
would suggest an Ansatz for $\overline{\psi}_{0}$ as a superposition
of modes in a generalized Fourier sin-series, as analyzed in Ref.
\cite{Boettcher17a}. However, the discussion in Appendix C confirms
that even the simplest Ansatz of considering merely the two closest
poles to $z=1$ suffices here, and we may write
\begin{eqnarray}
\psi_{0,t} & \sim & N^{\epsilon}\sin\left(\theta_{k}t\right)\frac{\psi_{IC}}{\sqrt{N}},\label{eq:psi0t}
\end{eqnarray}
which, after Laplace transformation according to Eq.~(\ref{eq:LaplaceT}),
produces two Laplace-poles at $z_{0}=e^{\pm i\theta_{k}}$ symmetrically
impinging on $z=1$ along the unit-circle in the complex-$z$ plane:
\begin{eqnarray}
\overline{\psi}_{0}(z) & \sim & \frac{N^{\epsilon}}{2i}\left(\frac{1}{1-ze^{i\theta_{k}}}-\frac{1}{1-ze^{-i\theta_{k}}}\right)\frac{\psi_{IC}}{\sqrt{N}},\nonumber \\
 & \sim & N^{\epsilon}\left[\frac{1}{\theta_{k}}\zeta^{0}-\frac{1}{\theta_{k}^{3}}\zeta^{2}+\frac{1}{\theta_{k}^{3}}\zeta^{3}+\frac{1}{\theta_{k}^{5}}\zeta^{4}+\ldots\right]\frac{\psi_{IC}}{\sqrt{N}}.\label{eq:psi0poles}
\end{eqnarray}
Then, we match Eqs.~(\ref{eq:psi0zeta}) and\ (\ref{eq:psi0poles})
term-by-term in $\zeta$ to get 
\begin{eqnarray}
\frac{N^{\epsilon}}{\theta_{k}}\sim\lambda_{1}^{k}, & \qquad & \frac{N^{\epsilon}}{\theta_{k}^{3}}\sim\lambda_{1}^{2k}\lambda_{2}^{k},\label{eq:Stheta}
\end{eqnarray}
which provides for the characteristic period and the amplitude at
time $t_{{\rm opt}}$ with $\sin\left(\theta_{k}t_{{\rm opt}}\right)^{2}=1$:
\begin{eqnarray}
T(N) & \sim & \frac{1}{\theta_{k}}\sim\sqrt{\lambda_{1}^{k}\lambda_{2}^{k}}\sim N^{\frac{d_{w}^{Q}}{d_{f}}},\label{eq:mainresult}\\
\left|\psi_{0,t}\right|^{2} & \sim & \left(\frac{N^{\epsilon}}{\sqrt{N}}\right)^{2}\sim\frac{\lambda_{1}^{k}}{N\lambda_{2}^{k}}\sim N^{1-2\frac{d_{w}^{Q}}{d_{f}}},\nonumber 
\end{eqnarray}
where we have identified the eigenvalues with the appropriate dimensions
as given in Eq.~(\ref{eq:dfw}). In fact, we have extended the RG-expansion
in Eq.\ (\ref{eq:psi0zeta}) to two more orders and found that they
scale consistently with the $\zeta^{3}$ and $\zeta^{4}$-terms of
Eq.~(\ref{eq:psi0poles}). In Figs.\ \ref{fig:SearchCollapseDSG},
we demonstrate that the scaling in Eq.~(\ref{eq:mainresult}) perfectly
collapses the data we have obtained from numerical simulations of
quantum search on DSG. They yield the computational complexity stated
in Eq.\ (\ref{eq:costNoTulsi}) and the naive scaling shown in Fig.\ \ref{fig:cplot}. 

In fact, those values for the rescaling of $p=\left|\psi_{0,t}^{2}\right|$
and $T$ in Eq.\ (\ref{eq:mainresult}) have been studied numerically
before by Patel and Raghunathan\ \cite{Patel12}, who found $p\sim N^{\text{\textminus}0.440(4)}$
and $T\sim N^{0.730(2)}$ for a coined quantum search on a regular
Sierpinski lattice, which is not too far from the analytical prediction
here: $2\frac{d_{w}^{Q}}{d_{f}}-1=\log_{3}5-1=0.464974\ldots$ and
$\frac{d_{w}^{Q}}{d_{f}}=\frac{1}{2}\log_{3}5=0.732487\ldots$. Recently,
Tamegai et al \cite{Tamegai18}) found equivalent results also for
the Sierpinski carpet (which is not renormalizable). Similarly, Marquezino
et al.\ \cite{Marquezino12} simulated a quantum search with a modified
Grover coin on the Hanoi network (HN3) and found $p\sim N^{-0.37}$
and $T\sim N^{0.65}$, in reasonable agreement with the analytical
prediction of $2\frac{d_{w}^{Q}}{d_{f}}-1=0.30576\ldots$ and $\frac{d_{w}^{Q}}{d_{f}}=0.652879\ldots$,
using $d_{w}^{Q}=2-\log_{2}\frac{\sqrt{5}+1}{2}$ and $d_{f}=2$ found
for this network\ \cite{Boettcher14b}. Both of these numerical studies
also considered successful implementations of Tulsi's method to optimize
the overlap to become $p\sim1$, which we explore analytically with
the RG in the following.

\subsubsection{Optimization with Tulsi's Method:\label{subsec:Extension-to-Tulsi's}}

Tulsi\ \cite{Tul08} realized that the interplay between walk-operator
and search-operator in an implementation of Grover's algorithm on
a low-dimensional geometry can be further optimized by adding at most
two ancilla qubits\ \cite{tulsi2012}. Thereby, each is doubling
the dimensions to the internal coin-space of the quantum walk (which
has been compared to giving a Dirac-fermion a position-dependent mass\ \cite{Patel12}).
This minimal extension inserts a tunable parameter $\tau$ that allows
to ``buffer'' more weight $\left|\psi_{0,t}^{2}\right|$ only at
the sought-after site $w$ in just the right amount so as to optimize
$p=\left|\psi_{0,t}^{2}\right|$ to attain a finite, $N$-independent
value just at the time of measurement. The optimal choice for this
parameter itself does depend on $N$ but is independent of $w$. While
the implementation details are technical and have been deferred to
Appendix D, the calculation follows that in Sec.\ \ref{subsec:General-Considerations-for-DSG}
closely but with somewhat enlarged matrices. In the end, we obtain
relations almost identical to Eq.\ (\ref{eq:psi0zeta}) but with
an overall factor of $\cot\tau$. Then, Eq.\ (\ref{eq:Stheta}) generalizes
to:
\begin{eqnarray}
\frac{N^{\epsilon}}{\theta_{k}}\sim\lambda_{1}^{k}\cot\tau, & \qquad & \frac{N^{\epsilon}}{\theta_{k}^{3}}\sim\lambda_{1}^{2k}\lambda_{2}^{k}\cot\tau.\label{eq:SthetaTulsi}
\end{eqnarray}
Note that the limit $\tau\to0$, in which the part of the product-space
linked by Tulsi's ancilla qubits would disconnect, emerges as a singular
limit, $\cot\tau\to\infty$, in the RG. Taking the ratio of both expressions
in Eq.\ (\ref{eq:SthetaTulsi}) cancels the $\tau$-dependence, signifying
that the quantum transport scaling expressed by $T\sim1/\theta_{k}$
found in Eq.\ (\ref{eq:mainresult}) remains unaffected, consistent
with the fact that the ancilla merely acts only locally at site $w$.
However, the amplitude at site $w$, obtained by the product of both
relations in Eq.\ (\ref{eq:SthetaTulsi}) now becomes 
\begin{eqnarray}
\left|\psi_{0,t}\right|^{2} & \sim & \frac{\lambda_{1}^{k}}{N\lambda_{2}^{k}}\cot^{2}\tau,\label{eq:Tulsi_p}
\end{eqnarray}
which in reference to Eq.\ (\ref{eq:mainresult}) we are free to
optimize via 
\begin{equation}
\tau\sim N^{\frac{d_{w}^{Q}}{d_{f}}-\frac{1}{2}}\ll1,\label{eq:Tulsi_opt}
\end{equation}
such that $p=\left|\psi_{0,t}^{2}\right|\sim1$, mindful of the fact
that $p$ is bounded by unity, of course. This analytical results
reproduces again the numerical predictions and the scaling relations
found\ \cite{Patel12,Marquezino12} for Tulsi's parameter $\tau$.

\subsection{Discussion of Search Operator $D=\mathbb{I}$\label{subsec:Discussion-of-D=00003DI}}

\begin{figure}
\includegraphics[viewport=70bp 0bp 600bp 720bp,clip,angle=270,width=0.48\columnwidth]{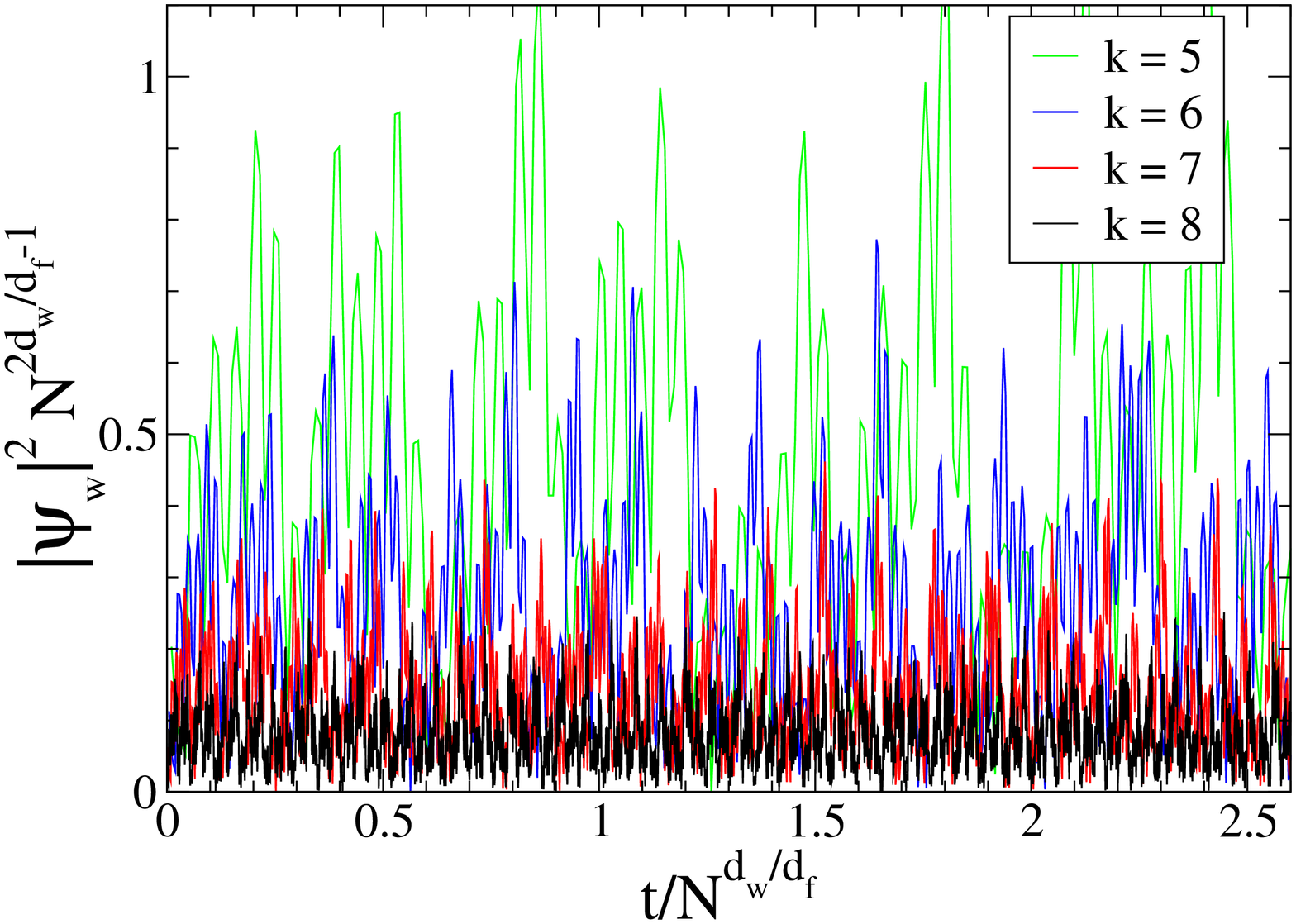}\hfill{}\includegraphics[viewport=70bp 0bp 600bp 720bp,clip,angle=270,width=0.48\columnwidth]{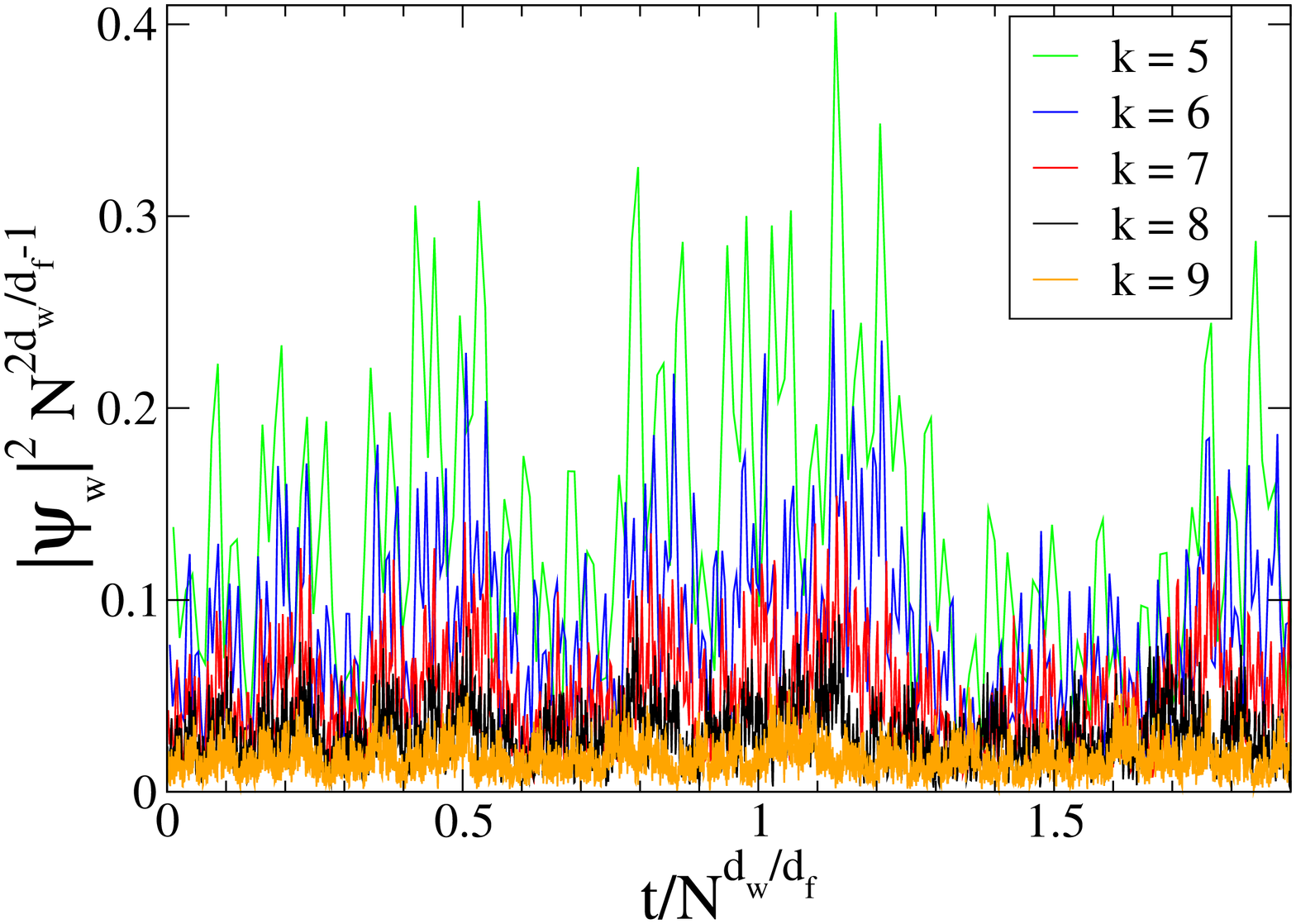}

\caption{\label{fig:NonSearchCollapse}Plot of the probability $p=\left|\psi_{w,t}^{2}\right|$
to detect the quantum walk at site $w=0$ as a function of time $t$
for DSG of size $N_{k}=3^{k}$ with defective search operators, $D=\mathbb{I}$
(top) and $D$ non-reflective (bottom). Although we have plotted the
simulation data on the same scale as in Fig.\ \ref{fig:SearchCollapseDSG},
it is apparent that no effective search is achieved, as predicted
by the RG in Eq.\ (\ref{eq:psi0_D=00003DI}) and Eq.\ (\ref{eq:psi_tnonR}),
respectively. There are no discernible peaks, and the probability
$p$ to find anything is decaying despite the indicated rescaling
with size. Instead, we find $p\sim\frac{1}{N}$ throughout, equivalent
to a classical random search for both cases. }
\end{figure}

With the preceding methods, we can also address interesting questions
regarding the universality of the results. We have shown that the
search operator with the choice of $D$ in Eq.\ (\ref{eq:Dmatrix})
provides a scaling of the complexity that is independent of the parameter
$\gamma$. In turn, we find that $D=\mathbb{I}$, another choice that
satisfies the conditions on the search operator in Eq.\ (\ref{eq:Dcondition}),
will not allow to accumulate weight at the sought-after site $w$.
Following Eq.\ (\ref{eq:G_i}) in Sec.\ \ref{subsec:General-Considerations-for-DSG},
$\mathbb{I}-2G\left(z\right)D$ in Eq.~(\ref{eq:DSGPsi0}) is again
singular at order $\zeta^{-1}$, yet, even its inverse in Eq.~(\ref{eq:DSGPsi0})
possesses a leading contribution of order $\zeta^{-1}$ and has the
expansion:
\begin{eqnarray*}
\left[\mathbb{I}-2G\left(z\right)D\right]^{-1} & \sim & -\frac{2}{9\left({\cal A}\lambda_{1}^{k}\right)}X_{-1}\zeta^{-1}+X_{0}\zeta^{0}\\
 &  & \quad-\frac{5\left({\cal B}\lambda_{2}^{k}\right)}{12}X_{1}\zeta^{1}+\ldots,
\end{eqnarray*}
with 
\begin{eqnarray*}
X_{0} & = & \frac{1+\sin(\eta)}{\cos(\eta)}\left[\begin{array}{cc}
0 & 1\\
-1 & 0
\end{array}\right],\\
X_{i\not=0} & = & \left[\begin{array}{cc}
1 & -\frac{1+\sin\eta}{\cos\eta}\\
-\frac{1+\sin\eta}{\cos\eta} & \left(\frac{1+\sin\eta}{\cos\eta}\right)^{2}
\end{array}\right].
\end{eqnarray*}
Amazingly, for all $i\not=0$ the matrices $X_{i}$ are identical
in each order of $\zeta^{i}$ for large $k$. Now, $F$ in Eq.\ (\ref{eq:Fz})
annihilates all such $X_{i}$, i.e., $X_{i\not=0}F\equiv0$. Thus,
the combination $\left[\mathbb{I}-2G\left(z\right)D\right]^{-1}F\left(z\right)$
in Eq.~(\ref{eq:DSGPsi0}) results in a single term, 
\begin{equation}
\overline{\psi}_{0}\sim\frac{1}{2\zeta}\left[\begin{array}{cc}
-1-\sin\eta & -\cos\eta\\
\frac{\left(1+\sin\eta\right)^{2}}{\cos\eta} & 1+\sin\eta
\end{array}\right]\frac{\psi_{IC}}{\sqrt{N}},\label{eq:psi0_D=00003DI}
\end{equation}
near $z=1$, entirely independent of $k$. Hence, it remains $\left|\psi_{0,t}^{2}\right|\sim\frac{1}{N}$
for all times. We show simulations for $\left|\psi_{0,t}^{2}\right|$
with $D=\mathbb{I}$ for various sizes $N$ in Fig.\ \ref{fig:NonSearchCollapse}.

\subsection{Discussion of Non-Reflective Search Operators\label{subsec:Discussion-of-Non-Reflective}}

In a further exploration of universality classes for quantum search,
we want to investigate the effect of more general search operators.
Tulsi\ \cite{tulsi2012} has shown that a search operator ${\cal R}_{w}$
which is non-reflective should not affect the complexity of quantum
search significantly. However, that discussion assumed that the network
was complete. As a simple test whether the reflectivity condition
on ${\cal R}_{w}$ can be relaxed, we generalize Eq.\ (\ref{eq:Dmatrix})
to
\begin{equation}
D(\phi,\gamma)=e^{i\phi}\cos\phi D(\gamma),\label{eq:DnonR}
\end{equation}
which satisfies Eq.\ (\ref{eq:Dcondition}) but is not hermitian,
so $D^{2}\not=D$ for all $\phi\not=0$. For such a case, we find
that the RG-analysis produces a very different result that dramatically
changes in the limit $\phi\to0$. 

With the matrix $D(\phi,\gamma)$, the combination $\mathbb{I}-2G\left(z\right)D$
in Eq.~(\ref{eq:DSGPsi0}) is also singular at order $\zeta^{-1}$,
and its inverse in Eq.~(\ref{eq:DSGPsi0}) has a similarly leading
contribution of order $\zeta^{0}$. However, the key cancelation that
brought $\lambda_{2}$ to prominence in the $\zeta^{1}$-term of $G(z)$
in Eq.\ (\ref{eq:Gzetaseries}) is undone in this inversion, due
to the non-reflectivity of $D$: At each order in $\zeta^{j}$, the
most divergent term in $k$ is always $\sim\left[\left(1-e^{i\phi}\cos\phi\right)\lambda_{1}^{k}\right]^{j}$,
making $\lambda_{2}$ irrelevant unless $\phi=0$. This property continuous
also for $\left[\mathbb{I}-2G\left(z\right)D\right]^{-1}F\left(z\right)$
in Eq.~(\ref{eq:DSGPsi0}), but incurring an overall factor of $\lambda_{1}^{k}$
as $F(z)$ again annihilates the leading term while providing a factor
of $\zeta^{-1}$. Leaving constants of unit-order aside, we then have
from Eq.\ (\ref{eq:DSGPsi0}):
\begin{eqnarray}
\overline{\psi}_{0} & \sim & \lambda_{1}^{k}\left\{ \sum_{j=0}^{\infty}\left[\left(1-e^{i\phi}\cos\phi\right)\lambda_{1}^{k}\zeta\right]^{j}\right\} \frac{\psi_{IC}}{\sqrt{N}},\nonumber \\
 & \sim & \frac{N}{1-\left(1-e^{i\phi}\cos\phi\right)N(z-1)}\,\frac{\psi_{IC}}{\sqrt{N}},\label{eq:psi_zDnonR}
\end{eqnarray}
since $\lambda_{1}^{k}=N$. The inverse Laplace transform then yields
\begin{eqnarray}
\psi_{0,t} & \sim & \exp\left\{ -\frac{t}{\left(e^{i\phi}\cos\phi-1\right)N}\right\} \frac{\psi_{IC}}{\sqrt{N}}.\label{eq:psi_tnonR}
\end{eqnarray}
Ignoring the (rather approximate) complex exponential, which represents
a more general function that is bounded for all times $t$, Eq.\ (\ref{eq:psi_tnonR})
again suggest that $p$ will not exceed classical scaling, $\sim\frac{1}{N}$.
We show simulations for $\left|\psi_{0,t}^{2}\right|$ with $\phi=\frac{\pi}{4}$
for various sizes $N$ also in Fig.\ \ref{fig:NonSearchCollapse},
which confirms the RG-prediction.

\section{Discussion\label{sec:Discussion}}

We have indications to believe that the bounds in Eqs.~(\ref{eq:costNoTulsi}-\ref{eq:CostTulsi})
are generic for any network characterized in terms of the dimensions
$d_{w}^{Q}$ and $d_{f}$, or $d_{s}$, as depicted in Fig.\ \ref{fig:cplot}.
It is straightforward to extend this calculation to other networks,
such as the networks MK3 and MK4 discussed in Ref.~\cite{Boettcher14b},
which lead to identical conclusions aside from minor details in the
analysis~\cite{Boettcher17a}. A similar RG-analysis has been applied
previously to continuous-time quantum search algorithms~\cite{LiBo16}.
Since many quantum computing tasks are similarly defined over a network
geometry of interacting variables, we anticipate that our findings
would inspire equivalent studies for a broad range of quantum algorithms
in the future. For instance, quantum walks also drive the leading
quantum algorithm for the element distinctness problem~\cite{Ambainis07},
for finding graph isomorphisms~\cite{Rudinger13}, as well as for
other decision-making processes~\cite{Farhi1998}.

\paragraph*{Acknowledgements:}

SB acknowledges financial support from CNPq through the ``Ciência
sem Fronteiras'' program and thanks LNCC for its hospitality. RP
acknowledges financial support from Faperj and CNPq.

\bibliographystyle{apsrev4-1}
\bibliography{/Users/stb/Boettcher}

\section*{Appendix\label{sec:Appendix}}

\subsection{Generalized Unitarity Conditions for Quantum Walks on DSG\label{subsec:Universality-in-quantum}}

\begin{figure}[b!]
\hfill{}\includegraphics[viewport=50bp 40bp 320bp 390bp,clip,width=0.3\columnwidth]{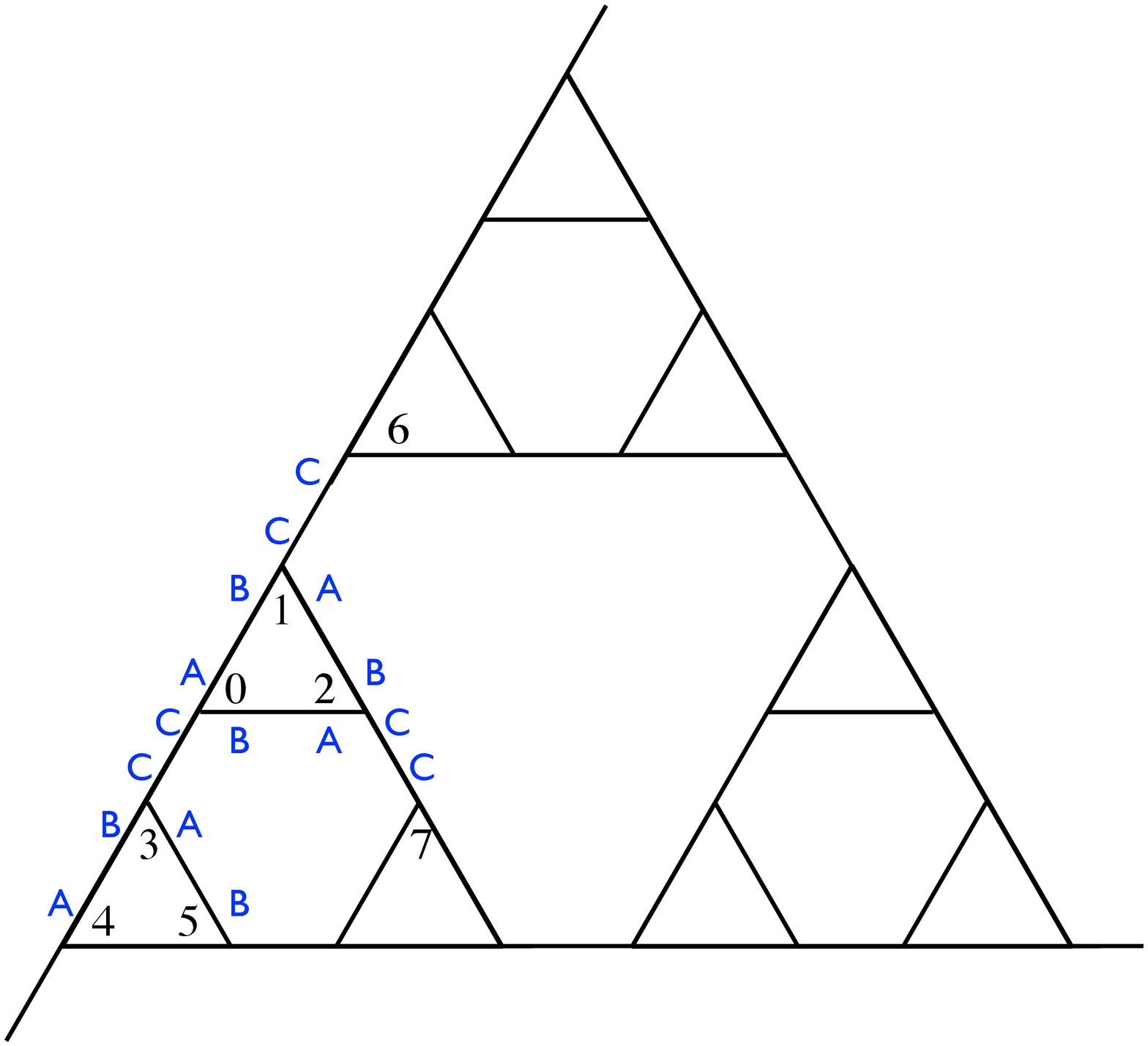}\hfill{}

\caption{\label{fig:DSG_UnitarityMask}A generic site $0$ and seven sites
$1,\ldots,7$ that are at most two hops away from $0$ on a dual Sierpinski
gasket (DSG). Only the relevant hopping operators $A,B,C$ for Eq.\ (\ref{eq:U0})
are labeled here. }
\end{figure}

Here, we establish generalized unitarity conditions on the propagator
${\cal U}$ in the Master equation\ (\ref{eq:MasterEq}) for the
DSG network. For the terms of the propagator pertaining to a generic
site $x=0$ in the DSG, see Fig.\ \ref{fig:DSG_UnitarityMask}, we
find 
\begin{eqnarray}
{\cal U}_{0} & = & M\left(\left|0\left\rangle \right\langle 0\right|+\left|1\left\rangle \right\langle 1\right|+\left|2\left\rangle \right\langle 2\right|+\left|3\left\rangle \right\langle 3\right|\right)\nonumber \\
 &  & \quad+A\left(\left|0\left\rangle \right\langle 2\right|+\left|3\left\rangle \right\langle 4\right|+\left|2\left\rangle \right\langle 1\right|+\left|1\left\rangle \right\langle 0\right|\right)\label{eq:U0}\\
 &  & \quad+B\left(\left|0\left\rangle \right\langle 1\right|+\left|3\left\rangle \right\langle 5\right|+\left|2\left\rangle \right\langle 0\right|+\left|1\left\rangle \right\langle 2\right|\right)\nonumber \\
 &  & \quad+C\left(\left|0\left\rangle \right\langle 3\right|+\left|3\left\rangle \right\langle 0\right|+\left|2\left\rangle \right\langle 7\right|+\left|1\left\rangle \right\langle 6\right|\right),\nonumber 
\end{eqnarray}
where sites labeled $x=1,\ldots,7$ are all at most two hops away
from $x=0$. However, even of those, we merely keep transition operators
$\left|i\left\rangle \right\langle j\right|$ for which (1) $j=0$
so that ${\cal U}_{0}\left|0\right\rangle \not=0$, or (2) $i$ is
at most one hop away from $x=0$ (here, $i=0,1,2,3$). These are the
\emph{only} terms that can impact the unitarity condition applicable
to site $x=0$, i.e., 
\begin{eqnarray}
{\cal U}_{0}^{\dagger}{\cal U}_{0}\left|0\right\rangle  & = & {\cal U}_{0}^{\dagger}\left(M\left|0\right\rangle +A\left|1\right\rangle +B\left|2\right\rangle +C\left|3\right\rangle \right),\nonumber \\
 & = & \left(A^{\dagger}A+B^{\dagger}B+C^{\dagger}C+M^{\dagger}M\right)\left|0\right\rangle \nonumber \\
 &  & \quad+\left(A^{\dagger}B+B^{\dagger}M+M^{\dagger}A\right)\left|1\right\rangle \label{eq:UdaggerU0}\\
 &  & \quad+\left(A^{\dagger}M+B^{\dagger}A+M^{\dagger}B\right)\left|2\right\rangle \nonumber \\
 &  & \quad+\left(C^{\dagger}M+M^{\dagger}C\right)\left|3\right\rangle +A^{\dagger}C\left|4\right\rangle \nonumber \\
 &  & \quad+B^{\dagger}C\left|5\right\rangle +C^{\dagger}A\left|6\right\rangle +C^{\dagger}B\left|7\right\rangle .\nonumber 
\end{eqnarray}
As $\left|0\right\rangle $ is a \emph{generic} site, its unitarity,
$\left\langle x\right|{\cal U}_{0}^{\dagger}{\cal U}_{0}\left|0\right\rangle =\delta_{x,0}$,
obtained from Eq.~(\ref{eq:UdaggerU0}), then implies ${\cal U}_{0}^{\dagger}{\cal U}_{0}=\mathbb{I}$
for \emph{every} site with the constraints finally summarized in Eq.\ (\ref{eq:UnitarityDSG}).

\subsection{Renormalization Group (RG)\label{subsec:Renormalization-Group-(RG)}}

To accomplish the decimation of the sites $\overline{\psi}_{\left\{ 3,\ldots,8\right\} }$,
as indicated in Fig.~\ref{fig:DSGRGstep}, we need to solve the linear
system in Eqs.~(\ref{eq:DSG_master}) for $\overline{\psi}_{\left\{ 0,1,2\right\} }$.
(Note that the following procedure is equivalent to that in Ref.\ \cite{QWNComms13},
but significantly simplified by the assumption of symmetry, $A=B$,
among the hopping operators.) Thus, we expect that $\overline{\psi}_{\left\{ 3,\ldots,8\right\} }$
can be expressed as (appropriately symmetrized) linear combinations
\begin{eqnarray}
\overline{\psi}_{\left\{ 3,4\right\} } & = & P_{0}\overline{\psi}_{0}+Q\overline{\psi}_{\left\{ 1,2\right\} }+R\overline{\psi}_{\left\{ 2,1\right\} }+J\psi_{IC},\nonumber \\
\overline{\psi}_{\left\{ 5,8\right\} } & = & R_{0}\overline{\psi}_{0}+P\overline{\psi}_{\left\{ 1,2\right\} }+Q\overline{\psi}_{\left\{ 2,1\right\} }+J\psi_{IC},\label{eq:Ansatz}\\
\overline{\psi}_{\left\{ 6,7\right\} } & = & Q_{0}\overline{\psi}_{0}+P\overline{\psi}_{\left\{ 1,2\right\} }+R\overline{\psi}_{\left\{ 2,1\right\} }+J\psi_{IC}.\nonumber 
\end{eqnarray}
Inserting this Ansatz into Eqs.~(\ref{eq:DSG_master}) and comparing
coefficients provides consistently for the unknown matrices $\left\{ P,Q,R,J\right\} $:
\begin{eqnarray}
P & = & \left(M+A\right)P+A+CR,\nonumber \\
Q & = & \left(M+C\right)Q+AR,\nonumber \\
R & = & MR+AQ+CP,\label{eq:PQRJ}\\
J & = & I+\left(M+A+C\right)J.\nonumber 
\end{eqnarray}
Abbreviating $S=\left(\mathbb{I}-M-C\right)^{-1}A$ and $T=\left(\mathbb{I}-M-AS\right)^{-1}C$,
Eqs.~(\ref{eq:PQRJ}) have the solution: 
\begin{eqnarray}
P & = & \left(\mathbb{I}-M-A-CT\right)^{-1}A,\nonumber \\
R & = & TP,\nonumber \\
Q & = & SR,\label{eq:solPQRJ}\\
J & = & \left(\mathbb{I}-M-A-C\right)^{-1}I.\nonumber 
\end{eqnarray}

Finally, after $\overline{\psi}_{\left\{ 3,\ldots,8\right\} }$ have
been eliminated, we find 
\begin{eqnarray}
\overline{\psi}_{0} & = & \left(\left[M_{0}+2AP_{0}\right]+C_{0}\right)\overline{\psi}_{0}+A\left(Q+R\right)\left(\overline{\psi}_{1}+\overline{\psi}_{2}\right)\nonumber \\
 &  & \qquad+\left(I+2AJ\right)\psi_{IC},\label{eq:psi0RG}
\end{eqnarray}
and similar for $\overline{\psi}_{\left\{ 1,2\right\} }$. By comparing
coefficients between the renormalized expression in Eq.~(\ref{eq:psi0RG})
and the corresponding, \emph{self-similar} expression in the first
line of Eqs.~(\ref{eq:DSG_master}), we can identify the RG-recursions
\begin{eqnarray}
M_{k+1} & = & M_{k}+2A_{k}P_{k},\nonumber \\
A_{k+1} & = & A_{k}\left(Q_{k}+R_{k}\right),\nonumber \\
C_{k+1} & = & C_{k},\label{eq:RGrecur}\\
I_{k+1} & = & I_{k}+2A_{k}J_{k},\nonumber 
\end{eqnarray}
where the subscripts refer to $k$-renormalized (or, un-renormalized)
and $(k+1)$-renormalized form of the hopping operators. These recursions
evolve from the un-renormalized ($k=0$) hopping operators with 
\begin{eqnarray}
\left\{ M,A,C\right\} _{k=0} & = & z\left\{ M,A,C\right\} ,\nonumber \\
I_{k=0} & = & \mathbb{I}\,{\rm or}\,0.\label{eq:RGIC}
\end{eqnarray}
Note that the RG-recursion for $\left\{ M,A,C\right\} $, the ``engine''
that drives the walk dynamics, evolves \emph{irrespective} of the
specific problem under consideration and independently from $I_{k}$.
Only $I_{k=0}$ refers to the specific problem one may intend to study,
as we discuss in Sec.~\ref{sec:Methods}. Implementing these recursions
in \noun{Mathematica}, for example, allows a convenient and detailed
reproduction of the results presented in the main text. 

\subsection{Analysis considering many poles\label{subsec:Analysis-considering-many}}

Here, we present a more elaborate analysis of the Laplace-poles leading
to the main result in Eq.\ (\ref{eq:mainresult}). Instead of only
incorporating the poles closets to the real-$z$ axis, as in Eq.\ (\ref{eq:psi0t}),
we extend the discussion to allow for a diverging number of such poles,
as Fig.\ \ref{fig:DSGpoles} would suggest. Such a consideration
is well-advised and has proven necessary for some observables\ \cite{Boettcher17a},
although it will only serve to justify our approach in the main text
for the present case. 

Again, for $t>0$, $\psi_{0,t}$ should be a periodic function of
some fundamental period $T(N)=2\pi/\theta_{k}$, but now we want to
consider it as a generalized Fourier sin-series, to wit 
\begin{eqnarray}
\psi_{0,t} & \sim & N^{\epsilon}\left[\sum_{j=1}^{h(N)}f_{j}\,\sin\left(g_{j}j\theta_{k}t\right)\right]\frac{\psi_{IC}}{\sqrt{N}}.\label{eq:psi0t-1}
\end{eqnarray}
To see why this form is justified, we take the Laplace transform as
in Eq.~(\ref{eq:LaplaceT}) to find 
\begin{eqnarray}
\overline{\psi}_{0}(z) & \sim & N^{\epsilon}\left[\sum_{j=1}^{h(N)}\frac{f_{j}}{2i}\left(\frac{1}{1-ze^{ig_{j}j\theta_{k}}}-\frac{1}{1-ze^{-ig_{j}j\theta_{k}}}\right)\right]\frac{\psi_{IC}}{\sqrt{N}},\nonumber \\
 & \sim & N^{\epsilon}\left[\frac{S_{1}}{\theta_{k}}\zeta^{0}-\frac{S_{3}}{\theta_{k}^{3}}\zeta^{2}+\frac{S_{3}}{\theta_{k}^{3}}\zeta^{3}+\frac{S_{5}}{\theta_{k}^{5}}\zeta^{4}+\ldots\right]\frac{\psi_{IC}}{\sqrt{N}},\label{eq:psi0poles-1}
\end{eqnarray}
where we defined 
\begin{equation}
S_{m}=\sum_{j=1}^{h(N)}\frac{f_{j}}{g_{j}^{m}j^{m}}.\label{eq:Sm-1}
\end{equation}
The first line of Eq.~(\ref{eq:psi0poles-1}) reflects the observation,
shown in Fig.~\ref{fig:DSGpoles}, that $\overline{\psi}_{0}(z)$
possesses a set of Laplace-poles on the unit-circle in the complex-$z$
plane, symmetric around the real-$z$ axis, that increasingly impinge
on that real axis at $z=1$. Near there, these poles are roughly equally
spaced, as expressed by multiples of a phase-angle, $j\theta_{k}$,
where $g_{j}$ represents some almost-constant function of $j$ that
captures any irregularities in the spacings. The function $h(N)$
allows for the possibility that a diverging number of such poles could
contribute\ \cite{Boettcher17a}. In turn, the residues at those
poles, $N^{\epsilon}f_{j}$, are the amplitudes for each mode in Eq.~(\ref{eq:psi0t-1}).
As $\left|\psi_{0,t}\right|$ is bounded, so is $\left|f_{j}\right|N^{\epsilon-1/2}$
both as a function of index $j$ and $N$. Accordingly, there must
be some $m_{0}$ such that the sums in Eq.~(\ref{eq:Sm-1}) are convergent
for $m\geq m_{0}$, i.e., $S_{m\geq m_{0}}=O(1),$ independent of
$h(N)$. In fact, the boundedness of $f_{j}$ with $j$ implies that
$S_{m\geq2}=O(1)$. We find that the only consistent choice to match
the RG-results in Eq.~(\ref{eq:psi0zeta}) is to assume that also
$S_{1}$ is constant, hence, the number of poles that needs to be
considered, $h(N)$, does not impact the considerations. Then, we
match Eqs.~(\ref{eq:psi0zeta}) and (\ref{eq:psi0poles-1}) term-by-term
in $\zeta$ to get 
\begin{eqnarray}
N^{\epsilon}\frac{S_{1}}{\theta_{k}}\sim\lambda_{1}^{k}, & \qquad & N^{\epsilon}\frac{S_{3}}{\theta_{k}^{3}}\sim\lambda_{1}^{2k}\lambda_{2}^{k},\label{eq:Stheta-1}
\end{eqnarray}
which provides for the characteristic period and the amplitude factor
already shown in Eq.\ (\ref{eq:mainresult}).

\begin{figure}
\hfill{}\includegraphics[viewport=85bp 320bp 660bp 600bp,clip,width=1\columnwidth]{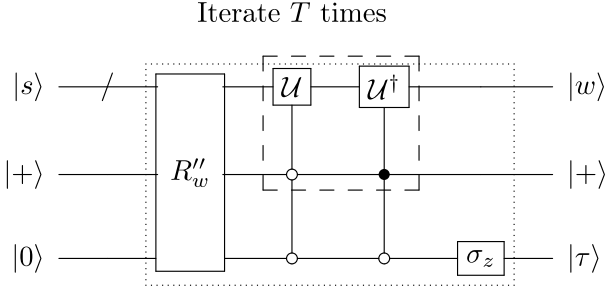}\hfill{}

\caption{\label{fig:TulsiDiagram}Diagram for the two-qubit extension of the
spatial Grover search in Tulsi's method. The inner (dashed) box describes
the action of the first qubit $H_{2}^{\prime}$ on the original walk-operator
${\cal U}$ that yields ${\cal U}^{\prime}$. It is only the second
qubit $H_{2}^{\prime\prime}$ that inserts the tunable parameter $\tau$,
whose optimized choice allows to evolve the state of the system from
its uniform initial state $\left|s\right\rangle $ to overlap with
probability $p\sim1$ with the sought-after state $\left|w\right\rangle $
after $T$ iterations, as determined in Eq.\ (\ref{eq:mainresult}),
of the search propagator ${\cal U}_{w}^{\prime\prime}={\cal U}^{\prime\prime}{\cal R}_{w}^{\prime\prime}$
(indicated by the faint outer box).}
\end{figure}

\subsection{Optimized search with Tusli's method\label{subsec:Tusli's-Method}}

Refs.\ \cite{Tul08,tulsi2012} outline an implementation of the spatial
Grover search algorithm for finite-dimensional networks that can dramatically
improve the probability to locate the sought-after site $w$ at the
optimal time for a measurement. While Tulsi introduces the idea first
to obtain the most efficient search algorithm to date on a square
lattice\ \cite{Tul08}, we follow here his generalization for arbitrary
unitary evolution operators\ \cite{tulsi2012}, such as ${\cal U}$
in Sec.\ \ref{subsec:The-Quantum-search}. Without further assumptions
on ${\cal U}$, we then require two extra qubits, as shown in the
diagram in Fig.\ \ref{fig:TulsiDiagram}. The Hilbert space then
becomes ${\cal H}^{\prime\prime}=H_{N}\otimes H_{{\cal C}}\otimes H_{2}^{\prime}\otimes H_{2}^{\prime\prime}$,
where ${\cal H}=H_{N}\otimes H_{{\cal C}}$ is the original Hilbert
space consisting of the real-space $H_{N}$ and the site-internal
coin-space $H_{{\cal C}}$. For example, the walk-operator ${\cal U}$
and the search-operator ${\cal R}_{w}=\mathbb{I}_{N}\otimes\mathbb{I}_{{\cal C}}-\left|w\right\rangle \left\langle w\right|\otimes\left(2D\right)$,
and the unitary ``search''-propagator ${\cal U}_{w}={\cal U}\cdot{\cal R}_{w}$,
as discussed in Sec.\ \ref{subsec:The-Quantum-search}, are operators
in ${\cal H}$. Then, let $H_{2}$ be a qubit 2-state space, in which
we conveniently define the projectors $\mathbb{P}_{s}=\left|s\right\rangle \left\langle s\right|$,
with $s\in\left\{ 0,1\right\} $ for each internal state of $H_{2}$.
Note that $\mathbb{P}_{0}+\mathbb{P}_{1}=\mathbb{I}_{2}$ and $\mathbb{P}_{0}-\mathbb{P}_{1}=\sigma_{z}$,
where $\sigma_{z}$ is a Pauli-matrix.

The first extension of the walk-operator with qubit $H_{2}^{\prime}$
entails (see diagram in Fig.\ \ref{fig:TulsiDiagram}):
\begin{eqnarray}
{\cal U}^{\prime} & = & \left(c_{1}^{\prime}{\cal U}^{\dagger}\right)\left(c_{0}^{\prime}{\cal U}\right),\nonumber \\
 & = & \left(\mathbb{I}_{N}\otimes\mathbb{I}_{{\cal C}}\otimes\mathbb{P}_{0}^{\prime}+{\cal U}^{\dagger}\otimes\mathbb{P}_{1}^{\prime}\right)\left({\cal U}\otimes\mathbb{P}_{0}^{\prime}+\mathbb{I}_{N}\otimes\mathbb{I}_{{\cal C}}\otimes\mathbb{P}_{1}^{\prime}\right),\nonumber \\
 & = & {\cal U}\otimes\mathbb{P}_{0}^{\prime}+{\cal U}^{\dagger}\otimes\mathbb{P}_{1}^{\prime}.\label{eq:U1prime}
\end{eqnarray}
Furthermore, for the target, we have
\begin{equation}
\left|w^{\prime}\right\rangle =\left|w\right\rangle \otimes\left|\gamma\right\rangle \otimes\left|+^{\prime}\right\rangle ,
\end{equation}
where in coin-space $\left|\gamma\right\rangle $ is such that we
get the operator $\left|\gamma\right\rangle \left\langle \gamma\right|=D\left(\gamma\right)$
in Eq.\ (\ref{eq:Dmatrix}), and where $\left|+\right\rangle =\left(\left|0\right\rangle +\left|1\right\rangle \right)/\sqrt{2}$.
Then, the search-operator ${\cal R}_{w}^{\prime}=\mathbb{I}_{N}\otimes\mathbb{I}_{{\cal C}}\otimes\mathbb{I}_{2}^{\prime}-2\left|w^{\prime}\right\rangle \left\langle w^{\prime}\right|$
and the search propagator ${\cal U}_{w}^{\prime}={\cal U}^{\prime}{\cal R}_{w}^{\prime}$
follow accordingly. (Under certain conditions on ${\cal U}$, this
first qubit may be redundant\ \cite{tulsi2012}.)

The second qubit $H_{2}^{\prime\prime}$ finally yields the walk-operator
\begin{eqnarray}
{\cal U}^{\prime\prime} & = & \left(\mathbb{I}_{N}\otimes\mathbb{I}_{{\cal C}}\otimes\mathbb{I}_{2}^{\prime}\otimes\sigma_{z}^{\prime\prime}\right)\left(c_{0}^{\prime\prime}{\cal U}^{\prime}\right),\nonumber \\
 & = & {\cal U}^{\prime}\otimes\mathbb{P}_{0}^{\prime\prime}-\mathbb{I}_{N}\otimes\mathbb{I}_{{\cal C}}\otimes\mathbb{I}_{2}^{\prime}\otimes\mathbb{P}_{1}^{\prime\prime},\label{eq:U2prime}
\end{eqnarray}
and target
\begin{equation}
\left|w^{\prime\prime}\right\rangle =\left|w^{\prime}\right\rangle \otimes\left|\tau^{\prime\prime}\right\rangle ,
\end{equation}
introducing the free parameter $\tau$ via
\begin{equation}
\left|\tau^{\prime\prime}\right\rangle =\sin\tau\left|0^{\prime\prime}\right\rangle +\cos\tau\left|1^{\prime\prime}\right\rangle .
\end{equation}
Then, we finally obtain the search propagator ${\cal U}_{w}^{\prime\prime}={\cal U}^{\prime\prime}{\cal R}_{w}^{\prime\prime}$
with the search operator
\begin{eqnarray*}
{\cal R}_{w}^{\prime\prime} & = & \mathbb{I}_{N}\otimes\mathbb{I}_{{\cal C}}\otimes\mathbb{I}_{2}^{\prime}\otimes\mathbb{I}_{2}^{\prime\prime}-2\left|w^{\prime\prime}\right\rangle \left\langle w^{\prime\prime}\right|,\\
 & = & \mathbb{I}_{N}\otimes\mathbb{I}_{{\cal C}}\otimes\mathbb{I}_{2}^{\prime}\otimes\mathbb{I}_{2}^{\prime\prime}\\
 &  & \quad-2\left|w\right\rangle \left\langle w\right|\otimes D(\gamma)\otimes D^{\prime}\left(\frac{\pi}{4}\right)\otimes D^{\prime\prime}(\pi-\tau),
\end{eqnarray*}
in an obvious adaptation of the matrix $D$ in Eq.\ (\ref{eq:Dmatrix}).

To follow the procedure outlined in Sec.\ \ref{subsec:The-Quantum-search},
we now merely need to first apply sequentially Eqs.\ (\ref{eq:U1prime})
and\ (\ref{eq:U2prime}) to each hopping operator $\left\{ M,A,C,I\right\} $
to obtain $\left\{ M^{\prime\prime},A^{\prime\prime},C^{\prime\prime},I^{\prime\prime}\right\} $.
While the entire fixed-point analysis of the RG in Sec.\ \ref{sec:Methods}
does not change, even in the search analysis in Sec.\ \ref{subsec:The-Quantum-search},
we only modify Eq.\ (\ref{eq:psi0zeta}) to read:
\begin{equation}
\overline{\psi}_{0}\sim\cot\tau\left\{ \left({\cal A}\lambda_{1}^{k}\right)F_{0}^{\prime\prime}\zeta^{0}+\left({\cal A}\lambda_{1}^{k}\right)^{2}\left({\cal B}\lambda_{2}^{k}\right)F_{2}^{\prime\prime}\zeta^{2}\right\} \frac{\psi_{IC}}{\sqrt{N}},\label{eq:psi0zetaTulsi}
\end{equation}
where the $F_{i}^{\prime\prime}$ are now the two-qubit enlarged versions
of those matrices in Eq.\ (\ref{eq:psi0zeta}). From this relation,
again in comparison with Eq.\ (\ref{eq:psi0poles}), follow the Tulsi-improved
Eqs.\ (\ref{eq:SthetaTulsi}) discussed in Sec.\ \ref{subsec:Extension-to-Tulsi's}.

\clearpage
\end{document}